\documentclass[12pt,preprint]{aastex}
\usepackage{psfig}
\usepackage{lscape}
\newcommand{\etal}{et al.~}

\shortauthors{Yao \& Wang}
\shorttitle{Absorption Line Spectroscopy of HISM}

\begin{document}

\title{X-ray Absorption Line Spectroscopy of the Galactic Hot 
Interstellar Medium}
\author{Yangsen Yao and Q. Daniel Wang}
\affil{Department of Astronomy, University of Massachusetts, 
  Amherst, MA 01003\\
  yaoys@astro.umass.edu and wqd@astro.umass.edu}

\begin{abstract}
We present an X-ray absorption line spectroscopic study of 
the large-scale hot interstellar medium (HISM) 
in the Galaxy. We detect Ne~IX K$_\alpha$ absorption lines 
in the {\sl Chandra} grating spectra of 
seven Galactic low-mass X-ray binaries. Three of these sources also show 
absorption of O~VII~K$_\alpha$, O~VII~K$_\beta$, 
and/or O~VIII~K$_\alpha$. Both centroid and width of the lines
are consistent with a Galactic HISM origin of the absorption. 
By jointly fitting the multiple lines, accounting for line saturation 
and assuming the collisional ionization equilibrium, 
we estimate the average absorbing gas temperature as 
$\sim 2.4\pm0.3\times10^6$ K (90\% confidence errors). 
We further characterize the spatial density distribution of the gas
as  ${\rm 6.4_{-1.7}^{+2.4}~exp(-|z|/1.2_{-0.5}^{+1.0}~kpc)
\times10^{-3}{\rm~cm^{-3}}} $ (a disk morphology)
or ${\rm 6.2_{-2.1}^{+3.8}~[ 1 + (R/2.3_{-1.1}^{+1.6}~kpc)^2]^{-1}
\times10^{-3} {\rm~cm^{-3}}}$ (a sphere morphology), 
where $z$ and $R$ are the distances from the Galactic plane and 
Galactic center (GC) respectively. 
Since nearly all the sight-lines with significant absorption lines detected
are somewhat toward GC and at low Galactic latitudes, 
these results could be severely biased.
More observations toward off-GC sight-lines and at high latitudes 
are urgently needed to further the study. Nevertheless,
the results demonstrate the excellent potential of X-ray absorption line 
spectroscopy  in the study of the HISM. 

\end{abstract}

\section{Introduction \label{sec:intro}}

It has long been theorized that a major, possibly dominant, phase of
the interstellar medium (ISM) in both the disk and the halo 
of the Galaxy is gaseous at temperature $T \sim 10^6$ K 
\citep{spi56, cox74, mck77, hei87, fer98}. The 
presence of this rarefied hot ISM (HISM) component affects the geometry and 
dynamics of the cool phases of the ISM, the propagation of cosmic rays and 
UV/soft X-ray photons,  
the Galactic disk-halo interaction, the distribution of 
metal abundances, and the feedback to the intergalactic medium. 
There is, however, no consensus on how much hot gas exists, how it is 
distributed in the Galaxy, and what thermal, chemical, and ionization
states it is in. 

Hot gas with temperatures of $T \gtrsim 10^6$ K may be
traced by X-rays. Indeed, it has been studied 
extensively with broad-band X-ray observations, producing
all-sky maps of the diffuse soft X-ray background 
(SXB, e.g., Snowden \etal 1997).
Furthermore, emission lines such as C~VI, O~VII, and O~VIII have 
been detected in a high spectral resolution observation made with 
microcalorimeters flown on a sounding rocket, confirming 
the thermal origin for much of the background \citep{mcc02}.
However, the X-ray emission 
carries little distance information, and its interpretation is typically 
subject to large uncertainties
in line-of-sight absorption. Therefore,
it is very difficult, if not impossible,
to infer the spatial and physical properties of the hot gas
from X-ray emission measurements alone. 

Alternatively, one can study the HISM by observing its
absorption against bright background X-ray sources. This capability is
now provided by the grating instruments aboard {\sl Chandra}  
and {\sl XMM-Newton} X-ray Observatories \citep{wei00, jan01}. 
Indeed, the highly-ionized oxygen/neon absorption lines
(O~VII, O~VIII, and/or Ne~IX), consistent with no velocity shift,  
have been detected in the spectra of 
several bright active galactic nuclei (AGNs):
PKS~2155--304 \citep{nic02, fang02}, 3C~273 \citep{fang03}, and 
MKN~421 \citep{nic03, ras03}. \citet{wang05}
have further detected narrow 
O~VII and Ne~IX absorption lines in the spectrum of LMC X--3; the 
equivalent widths (EWs) of these lines are similar to those seen in the 
AGN spectra, suggesting that the bulk of the absorbing material 
is within the 50 kpc distance of LMC~X--3. In addition,
 Futamoto \etal (2004) have detected O~VII and O~VIII
absorption lines in the sight-line toward a Galactic low
mass X-ray binary (LMXB) X1820--303;
the equivalent widths of the lines are substantially higher than 
those seen in the AGN spectra, consistent with a stronger diffuse X-ray 
emission observed in the Galactic bulge region, where this LMXB is located.

\begin{figure}[h]
  \centerline{
  \psfig{figure=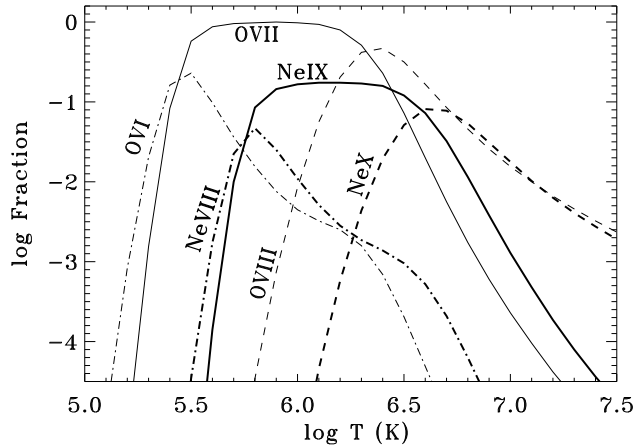,width=0.5\textwidth}}
  \caption{The ionization fractions of oxygen ({\sl thin lines}) 
    and neon (relative to all oxygen, {\sl thick lines}) as a function 
    of the CIE plasma temperature \citep{arn85}. 
    \label{fig:NeO_T} }    
\end{figure}

The detection of highly-ionized species in absorption lines
potentially provides a powerful tool in the study of the HISM.
For ease of reference, Fig.~\ref{fig:NeO_T} presents the relative ionization 
fraction of oxygen and neon as a function of temperature, for a plasma in a 
collisional ionization equilibrium (CIE) state. 
The X-ray absorption lines are sensitive to the gas over the entire
temperature range expected for the HISM.

\begin{itemize}
\item
  An individual absorption line alone, if resolved, gives a direct 
  measurement of the 
  kinematics and column density of the X-ray-absorbing gas,
  independent of the filling factor of the gas along the line of sight. 
\item
  Two or more absorption lines from the different transitions of the same ion, 
  i.e., K$_{\alpha}$, K$_{\beta}$, etc., even if not resolved, 
  may also be used to constrain  the kinematics and
  column density of the gas.
\item
  Two or more lines from the same element(s) but from different ionization 
  states may provide diagnostics of the thermal  
  and ionization states of the gas.
\item
  Multiple absorption lines of different elements (and transitions) 
  may further allow to measure their relative abundances.
\item
  If absorption lines
  are detected along many sight-lines and if the distances to the background 
  sources are known, one may then characterize the spatial distribution of 
  the gas. 
\item
  Because the absorption is relative to the local continuum level, 
  all such absorption line measurements are  
  independent of line-of-sight cool gas absorption! 
\item
  Of course, a joint analysis of the X-ray absorption line(s)
  with (line and/or broad-band) emission measurements may further give
  additional constraints on such important parameters as the average volume 
  density and the filling factor of hot gas.
\end{itemize}

We have carried out a systematic {\sl Chandra} archival study of the X-ray
absorption lines in the spectra of Galactic LMXBs. 
Because of their brightness and intrinsic spectral simplicity 
(relatively free from the complexity which could be caused by 
the stellar wind of a massive companion;
e.g., Schulz \& Brandt 2002), LMXBs are excellent background sources
for the study of the intervening X-ray-absorbing gas. We find that existing 
archival observations are 
already quite useful, although they were typically
not optimized for detecting absorption lines, in terms of both the instrument
setup and the exposure time. In addition to the work by \citet{fut04},
there have been a few reported X-ray absorption line studies that are
based on the grating observations of LMXBs. However, these studies
are focused on the absorption lines produced either by the cool phase of the 
ISM (e.g., Paerels \etal 2001; Juett \etal 2004) or by gas intrinsic to 
the binaries (e.g., Miller \etal 2004).

Here we report results from our archival study. We first describe our 
selection of the 
LMXBs and the {\sl Chandra} observations in \S\ref{sec:obs} and then introduce
our multiplicative absorption line model in \S\ref{sec:absline}. 
In \S\ref{sec:results}, 
we present the results of our absorption
line measurements. We discuss the origin (\S\ref{sec:wind}) and spatial
distribution (\S\ref{sec:spatial}) of the X-ray-absorbing gas and 
make comparisons with other relevant observations 
(e.g., O~VI and diffuse X-ray emission; \S\ref{sec:OVI}). Finally 
in \S\ref{sec:sum}. we summarize our results and conclusions.
Throughout the paper, errors are quoted at the 90\% confidence level.

\section{Source Selection and Observations \label{sec:obs}}

\begin{figure}[h]
  \centerline{
    \psfig{figure=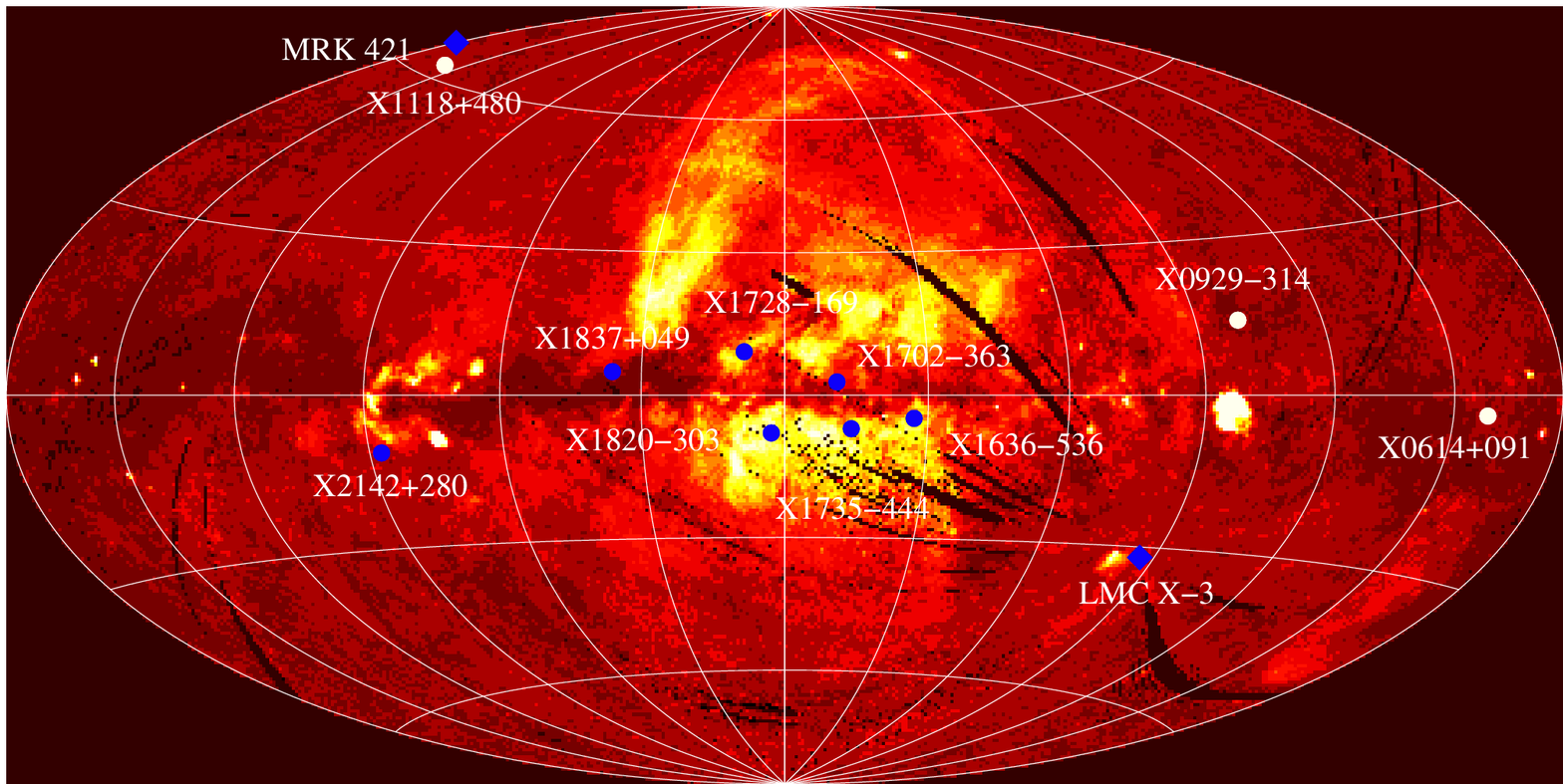,width=0.99\textwidth,angle=90}
  }
  \caption{Directions of X-ray sources in the present study, shown
    on a {\sl ROSAT} all-sky diffuse 3/4-keV-band intensity map 
    (in the Aitoff projection; Snowden et al. 1997). These sources are
    the 10 LMXBs ({\sl filled circle}) listed in Table~\ref{tab:sources},
    plus the two extragalactic sources, MRK~421 and LMC~X--3 
    ({\sl filled diamond}). The {\sl blue symbols} represent those with 
    significant absorption line detections, whereas the {\sl white symbols} 
    mark ones without a detection. \label{fig:galacticmap} }
\end{figure}

Our study use 17 {\sl Chandra} grating observations of 10 LMXBs,
available in the archive in April 2004 
(Table~\ref{tab:sources}; Fig.~\ref{fig:galacticmap}).
These sources are selected with two criteria:
1) a Galactic latitude $|b|\ge2^\circ$ to
minimize the effect of soft X-ray absorption by cool gas, and 2) 
a high signal-to-noise ratio $\gtrsim 7$ per 
bin at $\sim 0.9$ keV to have a good probability for 
a positive absorption line detection. 
We exclude those sources with 
absorption or emission lines that have been 
identified as intrinsic to the binaries (e.g., GX339--4, Miller \etal 2004; 
Her X-1, Jimenez-Garate \etal 2002). 

\begin{deluxetable}{lccccccc}
\tablewidth{0pt}
\tablecaption{Source Parameters\label{tab:sources}}
\tablehead{
       &            &  ($l$, $b$)     & $D$   & $z$   & $P$   & $A$         & \\
Source & Other Name & (deg)           & (kpc) & (kpc) & (hr)  & (R$_\odot$) & Ref.\tablenotemark{b}
}
\startdata
X1820--303 & NGC 6624   &   (2.79, -7.91)  & 7.6   &  1.0   & 0.19  & $<0.1$    & 1,2,3\\
X1728--169 & GX 9+9     &   (8.51, 9.04)   & 5.0   &  0.8   & 4.2   & 1.6       & 1,4,5\\
X1837+049  & Ser X--1   &   (36.12, 4.84)  & 8.4   &  0.7   & 13    & $\sim$3.4\tablenotemark{a}      & 1,4\\
X2142+380  & Cyg X--2   & (87.33, -11.22)  & 7.2   &  1.4   & 235   & 26.9      & 1,9,15\\
X1118+480  & $\cdots$   & (157.66, 62.32)  & 1.8   &  1.6   & 4.1   &$\cdots$   & 10, 11\\
X0614+091  & $\cdots$   & (200.88, -3.36)  &$\le3$ & $<0.2$ &$\le1$ &$\cdots$   & 12\\
X0929--314 & $\cdots$   & (260.11, 14.22)  &$\ge5$ &$\ge1.2$& 0.72  &$\cdots$   & 13,16\\
X1636--536 & V801 Ara   & (332.92, -4.82)  & 6.5   &  0.5   & 3.8   & $1.6\pm0.3$   & 1,4,7\\
X1735--444 & V926 Sco   & (346.05, -6.99)  & 7.1   &  0.9   & 4.6   & $1.75\pm0.02$ & 1,4,8\\
X1702--363 & GX 349+2   & (349.10, 2.75)   & 5.0   &  0.2   & 22    & $\sim$4.8\tablenotemark{a}      & 14, 4       
\enddata
\tablecomments{$D$, $z$, $P$, and $A$ are the distance, vertical distance 
away from the Galactic plane, orbit period, binary separation of the 
LMXBs.} 
\tablenotetext{a}{Estimated in this work by assuming that the 
companion is a main sequence star with a mass of $M_c\sim0.4$ M$_\odot$,
overflowing its Roche lobe and being accreted by a compact object 
with a mass of $M_X\sim1.4$ M$_\odot$ \citep{pac71}.}
\tablenotetext{b}{References: 
$^1$van Paradijs 1993; 
$^2$Kuulkers \etal 2003;
$^3$Stella \etal 1987;
$^4$Christian \& Swank 1997;
$^5$Schaefer 1990;
$^6$Homer \etal 1996;
$^7$Lawrence \etal 1983;
$^8$Smale \& Corbet 1991;
$^9$Cowley \etal 1979;
$^{10}$Cook \etal 2000;
$^{11}$McClintock \etal 2001a;
$^{12}$Brandt \etal 1992;
$^{13}$Markwardt \etal 2002;
$^{14}$Wachter \& Margon 1996;
$^{15}$Vrtilek \etal 2003;
$^{16}$Galloway \etal 2002;
}
\end{deluxetable}

\begin{figure}
  \centerline{
    \psfig{figure=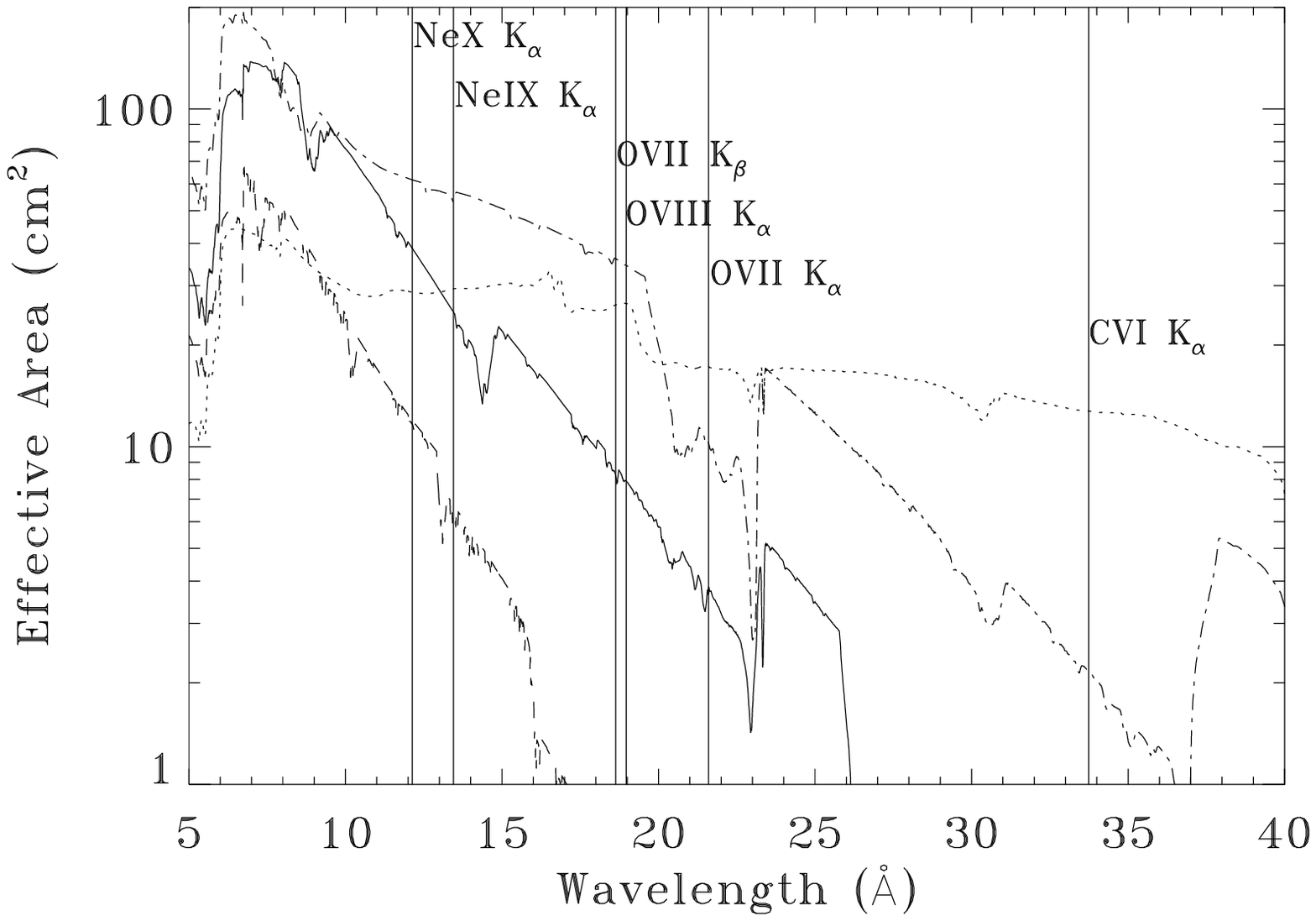,width=0.5\textwidth}
  }
  \caption{The effective area of the first-order spectra of the HEG 
    ({\sl dashed line}), 
    the MEG ({\sl solid line}), the LETG/ACIS ({\sl dash-dotted line}), and
    the LETG/HRC ({\sl dotted line}; with unsorted orders). The 
    {\sl vertical lines} mark the rest frame 
    wavelengths of highly ionized neon, oxygen, and carbon species 
    considered in this work. 
    The ACIS data were adopted from {\sl Chandra} cycle 4.
  \label{fig:eff}}   
\end{figure}

{\sl Chandra} carries two high spectral resolution instruments: 
the high energy transmission
grating (HETG, Markert \etal 1995; Canizares \etal 2000) and the 
low energy transmission grating 
(LETG, Pease \etal 2002). The HETG
consists of two grating assemblies, the high energy grating (HEG) and the medium 
energy grating (MEG). The LETG can be operated with either the 
advanced CCD imaging spectrometer (ACIS) or the high resolution camera (HRC),
whereas the HETG works only with the ACIS. The energy resolution of the
MEG, HEG, and LETG is 0.023, 0.012, and 0.05 \AA~(FWHM), respectively
\footnote{For detailed {\sl Chandra} instrumental information, 
please refer to Proposers' Observatory Guide
http://cxc.harvard.edu/proposer/POG/index.html}. 
For ease of reference, we
show in Fig.~\ref{fig:eff} the effective area of the telescope/instruments as 
a function of photon wavelength. 
With the expected range of hot gas column density, the most promising
absorption lines are at the
wavelengths of $\sim33$~\AA~(C~VI), 20~\AA~(O~VII and O~VIII), 
and 13~\AA~(Ne~IX and Ne~X). The rest-frame line wavelengths
are marked in Fig.~\ref{fig:eff}. Note that the effective area of the 
LETG/ACIS is small at $\sim33$~\AA, where the cool gas 
absorption of the continuum is also severe. 
Furthermore, for a LETG/HRC spectrum, the higher order confusion could be 
severe at this wavelength and sorting orders may also cause some potential 
uncertainties. Therefore, the C~VI is more difficult to detect 
than the other lines. The LETG/ACIS and LETG/HRC,
with an effective area of $\sim 20$/30 cm$^2$ at 20/13~\AA,  
should typically be the optimal combination
for detecting O~VII, O~VIII, and Ne~IX lines in one observation. 
The actual sensitivity 
of the LETG/ACIS also depends on the offset pointing, which determines
where the source is placed on the detector and whether 
the lines are detected in the front-illuminated CCDs or the 
back-illuminated CCDs. For HETG observations, only the data from the MEG 
are considered here because it is substantially more sensitive than the HEG.
Also because of the higher resolution of the MEG (by a factor of $\sim$ 2),
it can be more sensitive for detecting narrow Ne~IX and Ne~X lines than 
the LETG.

\begin{deluxetable}{llccc}
\tablewidth{0pt}
\tablecaption{{\sl Chandra} Observations\label{tab:observation}}
\tablehead{
           &       &  Obs. Date &              & Exposure \\
Source     & ObsID & (mm/dd/yy) & Grating/Detector & (ks) }
\startdata
X1820--303 & 98      & 03/10/00 & LETG/HRC  & 15.12 \\
           & 1021    & 07/21/01 & HETG/ACIS & 9.70  \\
           & 1022    & 09/12/01 & HETG/ACIS & 10.89 \\  
\hline
X1728--169 & 703     & 08/22/00 & HETG/ACIS & 22.44 \\
\hline       
X1837+049  & 700     & 06/13/00 & HETG/ACIS & 78.06 \\
\hline
X2142+380  & 87      & 04/24/00 & LETG/HRC  & 30.16 \\
           & 111     & 11/11/99 & LETG/ACIS & 6.2 \\ 
           & 1016    & 08/12/01 & HETG/ACIS & 15.13 \\
           & 1102    & 09/23/99 & HETG/ACIS & 28.98 \\
\hline        
X1118+480  & 1701    & 04/18/00 & LETG/ACIS & 27.83 \\
\hline
X0614+091  & 100     & 11/28/99 & LETG/HRC  & 26.23 \\
\hline        
X0929--314 & 3661    & 05/15/02 & LETG/ACIS & 17.96 \\
\hline         
X1636--536 & 105     & 10/20/99 & HETG/ACIS & 29.78 \\
           & 1939    & 03/28/01 & HETG/ACIS & 27.06 \\
\hline
X1735--444 & 704     & 06/09/00 & HETG/ACIS & 24.91 \\
\hline         
X1702--363 & 715     & 03/27/00 & HETG/ACIS & 10.98 \\
           & 3354    & 04/09/02 & HETG/ACIS & 35.21 \\
\hline                   
\enddata                 
\end{deluxetable}

Among the 17 observations, only six (on five
sources) used the LETG (Table~\ref{tab:observation}).
All of them, except for the short exposure on X2142+380 (ObsID 111, a  
calibration observation), have been analyzed for detecting absorption lines 
in previous studies. No significant highly ionized oxygen or neon 
absorption line is detected for X1118+480 \citep{mcc01b}, 
X0614+091 \citep{pae01}, and 
X0929--314 \citep{jue03}. Our re-analysis of the data confirms this result.
The detection of O~VII K$_\alpha$ and  
K$_\beta$, O~VIII K$_\alpha$, and 
Ne IX K$_\alpha$ absorption lines has been reported
for X1820--303, and an upper limit to the EW
of the O~VII absorption line has been set 
from the LETG spectrum of X2142+380 \citep{fut04}.
Our analysis of these two sources used both LETG and HETG observations
to achieve higher sensitivities. We report here for the first time
the detection of the highly ionized absorption lines 
toward X1728--169, X1837+049, X1636--546, X1735--444, and X1702--363. 

We re-process the observations, using the standard CIAO software 
(version 3.02) with the calibration database CALDB (version 2.25)
to extract source and background spectra. For each observation
we calculate the auxiliary response functions (ARFs) by running the
{\sl CIAO} thread {\sl fullgarf} for the positive and the negative
grating orders, and adopt the 
response matrix files (RMFs) from the CALDB directly. All the HETG/ACIS
observations had the zeroth order images either severely piled-up or
intentionally blocked to avoid telemetry saturation. We use the 
intersection of the two grating arms (both the HEG and the MEG) 
and the readout ``streak'' to
determine the position of the source in the detector 
and to calibrate the wavelength of the grating spectra
(e.g., Schulz \etal 2002). We then co-add
the positive and the negative order spectra  
to improve the counting statistics. 
We further combine the spectra from multiple observations of a
source, using the {\sl CIAO} thread {\sl add\_grating\_spectra}. 

\section{Implementation of an Absorption Line 
  Model \label{sec:absline} }

Published X-ray absorption line analyses typically use the following procedure
(e.g., Futamoto \etal 2004):
(1) adopting a smooth function (e.g., a power-law) to characterize the 
spectral continuum and an {\sl additive} negative {\it Gaussian} 
function to account for the absorption line profile; 
(2) fitting the observed spectrum to constrain the parameters of the functions;
(3) using the fitted EW of the line in a curve-of-growth analysis to 
estimate the ionic column density. This procedure is simple but typically
only adequate for individual unsaturated line analysis.  
The procedure also does not use all the information
available in the observed spectrum: details of the observed line shape and/or 
the intrinsic connections between multiple lines. Although   
a joint analysis of the ionic column densities inferred from 
the analysis of individual lines can, in principle,
be used to constrain the related physical parameters such as temperature
and ionic abundances, it is generally difficult to correctly propagate 
the non-Gaussian errors of the estimated parameters.
 
We have implemented a general model (referred here as {\it absline})
for absorption line fitting in the X-ray spectral analysis software 
package XSPEC. Following the description of 
the absorption process given by \citet{ryb79} and \citet{nic99}, 
the radiation transfer at photon energy $\varepsilon$ can be expressed as
\begin{equation} \label{equ:itau} 
  I(\varepsilon) = I_c(\varepsilon) e^{-\tau(\varepsilon)}, 
\end{equation}
and
\begin{equation} \label{equ:tau}
  \tau(\varepsilon) = N_i \frac{\pi e^2}{m_ec}f_{lu}\phi(\varepsilon),
\end{equation}
where $m_e$ is the electron mass, $c$ is the speed of light, and
$f_{lu}$ is the oscillator strength 
of the electron transition from a lower level to an upper level.
The ionic column density $N_i$ is a function of the reference element
(hydrogen as the default) column density $N_H$ and the plasma temperature $T$,
\begin{equation} \label{equ:ni}
  N_{i} = N_H f_{a} f_{C}(T),
\end{equation}
where $f_{a}$ is the corresponding element abundance   
relative to the reference element
and $f_{C}(T)$ is the ionic fraction as a function of $T$
(assuming that the absorbing gas is in a CIE state). 
The normalized Voigt function $\phi(\varepsilon)$ is a convolution 
of the intrinsic Lorentz line profile with a Doppler broadening (assumed to
be Gaussian):
\begin{eqnarray}
\phi(\varepsilon)& = & \int^{+\infty}_{-\infty}\frac{\gamma}{4\pi^2(\varepsilon-\varepsilon_l-\Delta\varepsilon)^2 + (\gamma/2)^2}
                \frac{e^{-(\Delta\varepsilon/\Delta\varepsilon_D)^2}}{\sqrt{\pi}\Delta\varepsilon_D}d(\Delta\varepsilon)\\
         & = &  \frac{1}{\sqrt{\pi}\Delta\varepsilon_D} H(a, u), \label{equ:phi}\\
 H(a, u) & = & \frac{a}{\pi} \int^{+\infty}_{-\infty}\frac{e^{-y^2}}{(u-y)^2 + a^2}dy~, \label{equ:hau}
\end{eqnarray}
where 
\begin{equation} \label{equ:au}
  a = \frac{\gamma}{4\pi\Delta\varepsilon_D},~~u = \frac{\varepsilon-\varepsilon_l}{\Delta\varepsilon_D}.
\end{equation}
Here $\gamma$ is the natural broadening damping factor, 
$\varepsilon_l$ is the systemic (rest frame if redshift 
$z$=0) energy of the line, and
$\Delta\varepsilon_D$ is the 
Doppler width
\begin{equation} 
  \Delta\varepsilon_D  =  \frac{b_v}{c} \varepsilon_l, 
\end{equation}
in which
\begin{equation}
  b_v =  \left(\frac{2kT}{m_i} + \xi^2\right)^{1/2}, \label{equ:b}
\end{equation}
where $m_i$ is the ionic mass, $\xi$ is the 
velocity dispersion due to extra-broadening (e.g., turbulence).
When $u$ is small (in the core region of a line), 
$H(a, u)\simeq e^{-u^2}$ gives a Gaussian profile, whereas when $u$ is 
large (in the wings of a line) $H(a, u)\simeq a/\pi u^2$ is close to a
Lorentz profile.

We define $e^{-\tau(\varepsilon)}$, part of Eq.~\ref{equ:itau},
as our multiplicative {\it absline} model, which is specified by 
five parameters: the plasma temperature $T$, central line energy 
$\varepsilon_l$, ion velocity dispersion $b_v$, 
the hot-phase hydrogen column density $N_H$ (or $N_i$ for a chosen ion), 
and the metal abundance $f_a$. 
Other parameters ($f_{lu}$ and $\gamma$) are given for 
a specific line (Table~\ref{tab:parameter}).
While $\varepsilon_l$ is very sensitive to the line centroid
in a fit, $b_v$ and $N_i$ can be constrained 
by the shape and intensity of the line. When multiple lines are present, 
one can conduct a joint fit, which may also allow for the estimation 
of $T$, $N_H$, and/or $f_a$ and their uncertainties. 

\begin{deluxetable}{lccc}
\tablewidth{0pt}
\tablecaption{Absorption Line Parameters\label{tab:parameter}}
\tablehead{
                 & rest $\lambda/\varepsilon_l$ &           & $\gamma$ \\
Line             & (\AA/eV)          & $f_{lu}$  & ($10^{-2}$ eV)  
}
\startdata
Ne~X~K$_\alpha$  & 12.134/1021.79        & 0.416     & 2.61  \\
Ne~IX~K$_\alpha$ & 13.448/921.95         & 0.657     & 3.35  \\
O~VII~K$_\beta$  & 18.629/665.55         & 0.146     & 0.39 \\
O~VIII~K$_\alpha$& 18.967/653.69         & 0.277     & 1.06 \\
O~VII~K$_\alpha$ & 21.602/573.95         & 0.696     & 1.37 \\
\enddata
\tablecomments{The $f_{lu}$ is the transition oscillator strength and $\gamma$ 
  is the natural broadening damping factor.
  The values of Ne~IX are adopted from \citet{beh02} and the others  
  from \citet{ver96}.} 
\end{deluxetable}

Note that in the {\sl absline} model, $b_v$ is used as an independent parameter 
(not constrained by Eq.~\ref{equ:b}), and can be smaller than the CIE 
thermal broadening $(2kT/m_i)^{1/2}$. Therefore, 
with the adopted Voigt function, the {\sl absline} model can be
applied as well to a photo-ionized plasma where $b_v\simeq0$.
The inclusion of the Lorentz profile in calculating the line profile
is important when the Doppler broadening is small (e.g., in an over-cooled 
or photo-ionized plasma) and/or when the column density is large. 
Using the Lorentz profile alone
gives a firm upper limit to the ionic column density (e.g.,  
Futamoto \etal 2004).

\begin{figure}[h]
\centerline{
  \psfig{figure=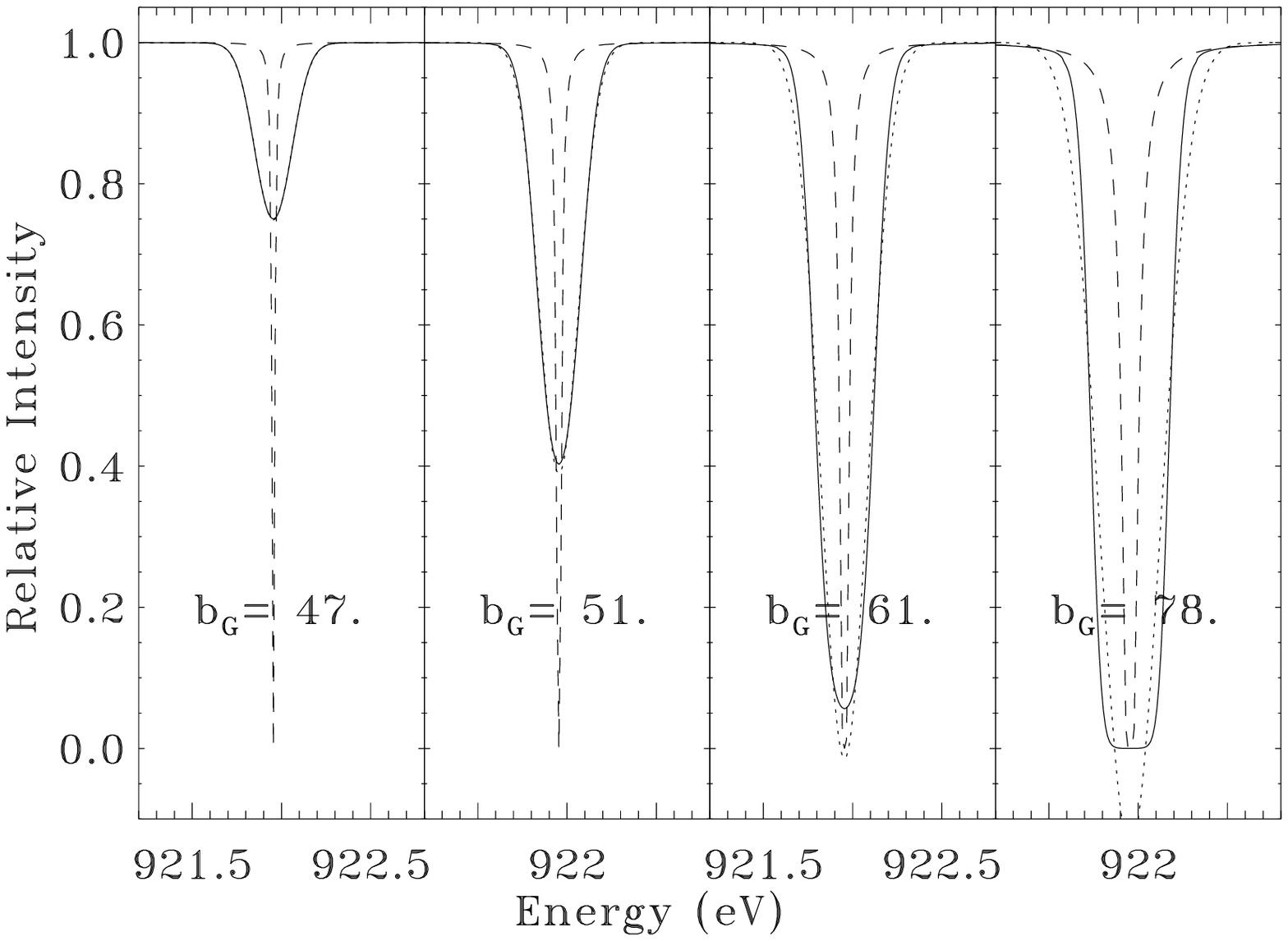,width=0.5\textwidth}
}
\caption{Comparison between the {\it absline} model of a Ne~IX~K$_\alpha$ 
  absorption line ({\sl solid line}) and the best-fit additive 
  {\it Gaussian} model ({\sl dotted}). The {\it absline} model assumes a 
  velocity dispersion 
  $b_v$=45 km~s$^{-1}$ (thermal velocity at T$\sim2.4\times10^6$ K) as well as
  ${\rm log(N_{NeIX})}$=15.0, 15.5, 16.0, and 16.5 cm$^{-2}$
  (for the panels from left to right), corresponding to optical depths of
  $\tau_0=$0.29, 0.91, 2.89, and 9.10 at the line centroid
  (921.95 eV $\sim$ 13.448 \AA). The $b_v=0$ Lorentz profiles 
  ({\sl dashed}) are also shown for reference
  in the panels. The limitations of the additive {\it Gaussian} model are
  characterized by the best fit $b_G$ values as they significantly differ from 
  $b_v$=45 km~s$^{-1}$. 
  \label{fig:abs-gau}}
\end{figure}

Fig.~\ref{fig:abs-gau} illustrates the difference between the additive 
{\it Gaussian} and the {\it absline} models.
The additive {\it Gaussian} model is a good
approximation to the {\it absline} model only when $\tau_0$ is 
small ($\lesssim 1$). When $\tau_0$ is large ($>1$), the additive
{\it Gaussian} profile deviates significantly from the {\it absline} profile 
and the inferred velocity dispersion $b_G$, for example, can be substantially
overestimated. 

\section{Analysis and Results \label{sec:results} }

Our absorption line analysis is based on the data in 12--14, 18--20, and 
20--22 \AA~bandpasses, which embrace the Ne~X~K$_\alpha$ and Ne~IX~K$_\alpha$, 
O~VII~K$_\beta$ and O~VIII~K$_\alpha$, and O~VII~K$_\alpha$ lines,
respectively (Table~\ref{tab:parameter}).
The chosen band width (2 \AA) is a compromise between maximizing
the counting statistics and reducing the effect of the potential deviation 
of the spectral continuum from our power-law characterization. 
The subtracted background contributes $\le5$\% to the source counts 
in the wavelength bandpasses for all sources. 
We apply both the commonly used additive {\it Gaussian} model and
the new {\it absline} model to the absorption line analysis of 
the 10 selected sources (Tables~\ref{tab:sources}-\ref{tab:observation}). 
The models, including both the continuum and the line(s), are convolved with the
instrument responses in XSPEC (version 11.3.1) before they were fitted to
the observations.
We choose the signal-to-continuum-noise ratio 
$S/N = 3\sigma$ (corresponding to a false detection
probability $\sim0.1\%$) as the detection threshold. 
Seven sources show significant 
Ne~IX~K$_\alpha$ absorption lines in the MEG spectra;
three of these sources also show O~VII~K$_\beta$, 
O~VIII~K$_\alpha$, and/or O~VII~K$_\alpha$ absorption line(s).
The O~VII~K$_\alpha$ absorption line is not significant 
in the LETG/HRC spectrum of X2142+380, which is consistent with
the previous work by \citet{fut04}, but we detect significant
Ne~IX~K$_\alpha$ and O~VII~K$_\beta$ absorption lines in the more
sensitive MEG spectra. All the detected lines are unresolved at
90\% confidence level. 
The lack of significant absorption lines in the spectra 
of X1118+480, X0614+091, and X0929--314 is consistent with 
the results from the previous studies as mentioned in \S\ref{sec:obs}.
No Ne~X~K$_\alpha$ absorption line is detected in any of the 10 sources.

\begin{figure}
  \centerline{ \hbox{
      \psfig{figure=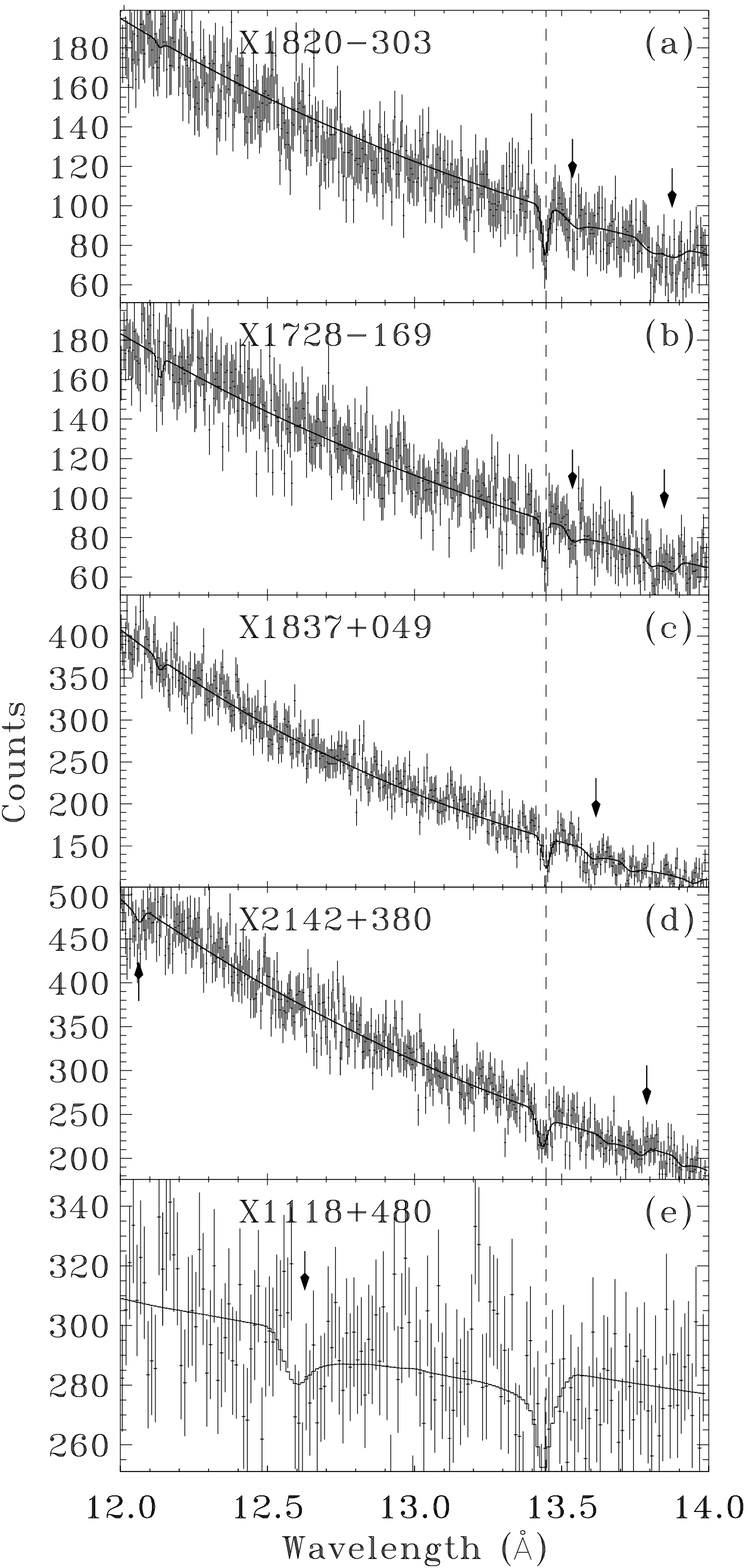,width=0.48\textwidth} 
      \psfig{figure=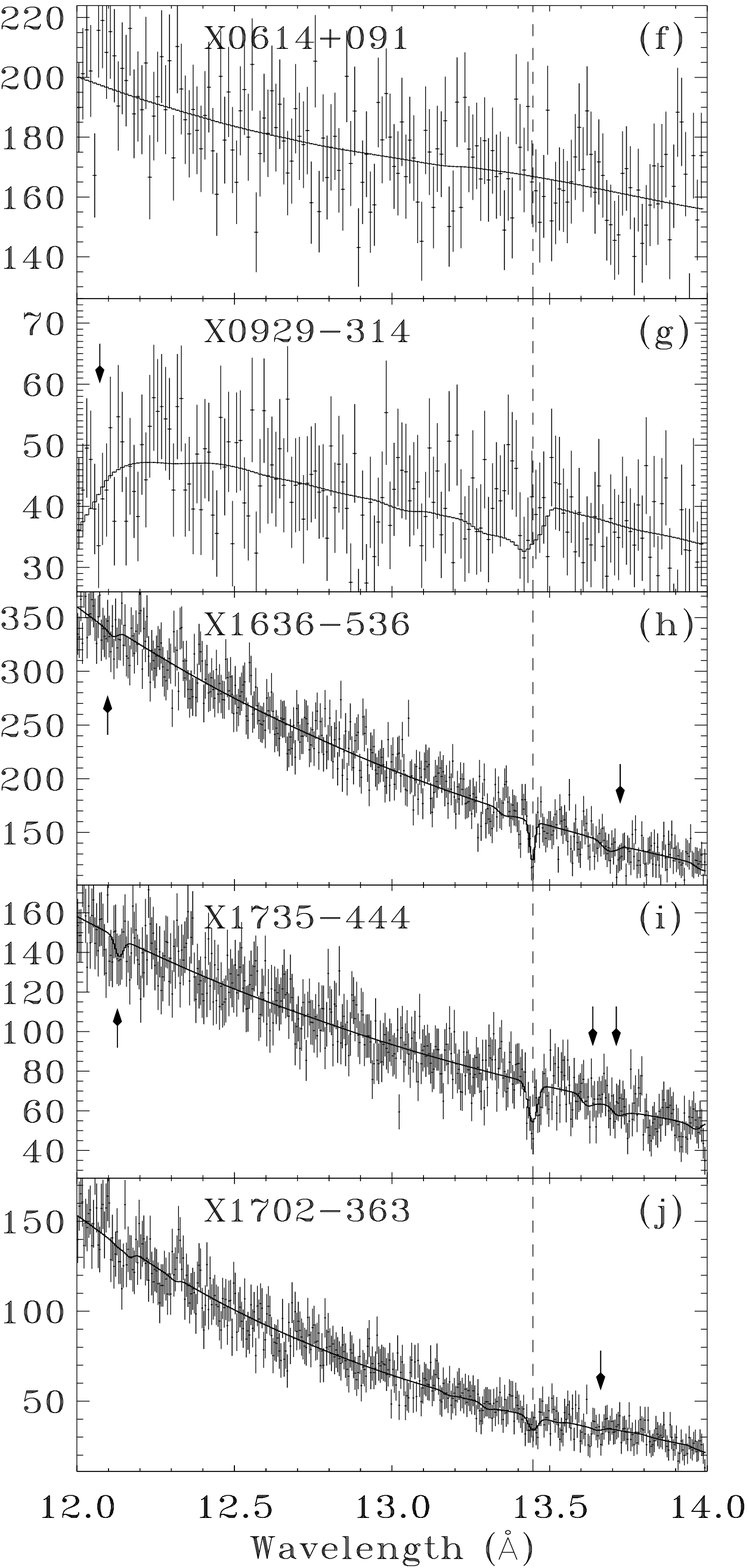,width=0.48\textwidth}
    }}
  \caption{The Ne~IX K$_\alpha$ absorption lines in the spectra of 
    the 10 sources. The {\sl solid line} in each panel represents a
    {\it Gaussian} plus 
    power-law model folded with the instrumental response; the vertical 
    {\sl dashed line} marks the rest frame wavelength of the line. 
    The apparent broad line features in {\sl (d)}, {\sl (e)}, 
    and {\sl (g)}, are partly due to dip in the instrument effective 
    area near the line central wavelength; the short vertical {\sl arrows}
    indicate other instrumental features.  
    The bin size is 12.5 m\AA~for LETG observations (panels {\sl e, f,} 
    {\sl g}) and is 5.0 m\AA~for HETG observations (the other panels).
    \label{fig:Neon} }
\end{figure}

For the detection of individual lines, it is convenient as a first pass 
to use the {\sl Gaussian} model. We estimate the line centroid energy and EW 
for each detected line. 
Fig.~\ref{fig:Neon} shows the {\sl Gaussian} model fits to the 
Ne~IX~K$_\alpha$ lines. For each sight-line,
we also estimate an upper limit to the EW of the 
Ne~X~K$_\alpha$ line by fixing its line centroid energy at the rest frame 
energy and jointly fitting its width with the 
Ne~IX~K$_\alpha$ line. 

Using the {\it absline} model, we jointly fit multiple lines of each source. 
Table~\ref{tab:fits} marks the included lines. In such 
a fit, we link the model parameters $b_v$, $T$,  and $N_H$ of the
lines; the slight dependence of $b_v$ on the different ion mass 
is neglected ($<5$ km~s$^{-1}$; see Eq.~\ref{equ:b}). 
Because of the limited number and quality of
the line detections, we use a fixed ISM abundance
\citep{wilms00} and assume the CIE for the X-ray-absorbing gas.
For those lines that are not detected, we 
fix $\varepsilon_l$ to their rest-frame energies.
Fig.~\ref{fig:X1820-1728} presents 
the velocity profiles of the jointly-fitted multiple lines.

\begin{figure}
  \centerline{
  \hbox{
    \psfig{figure=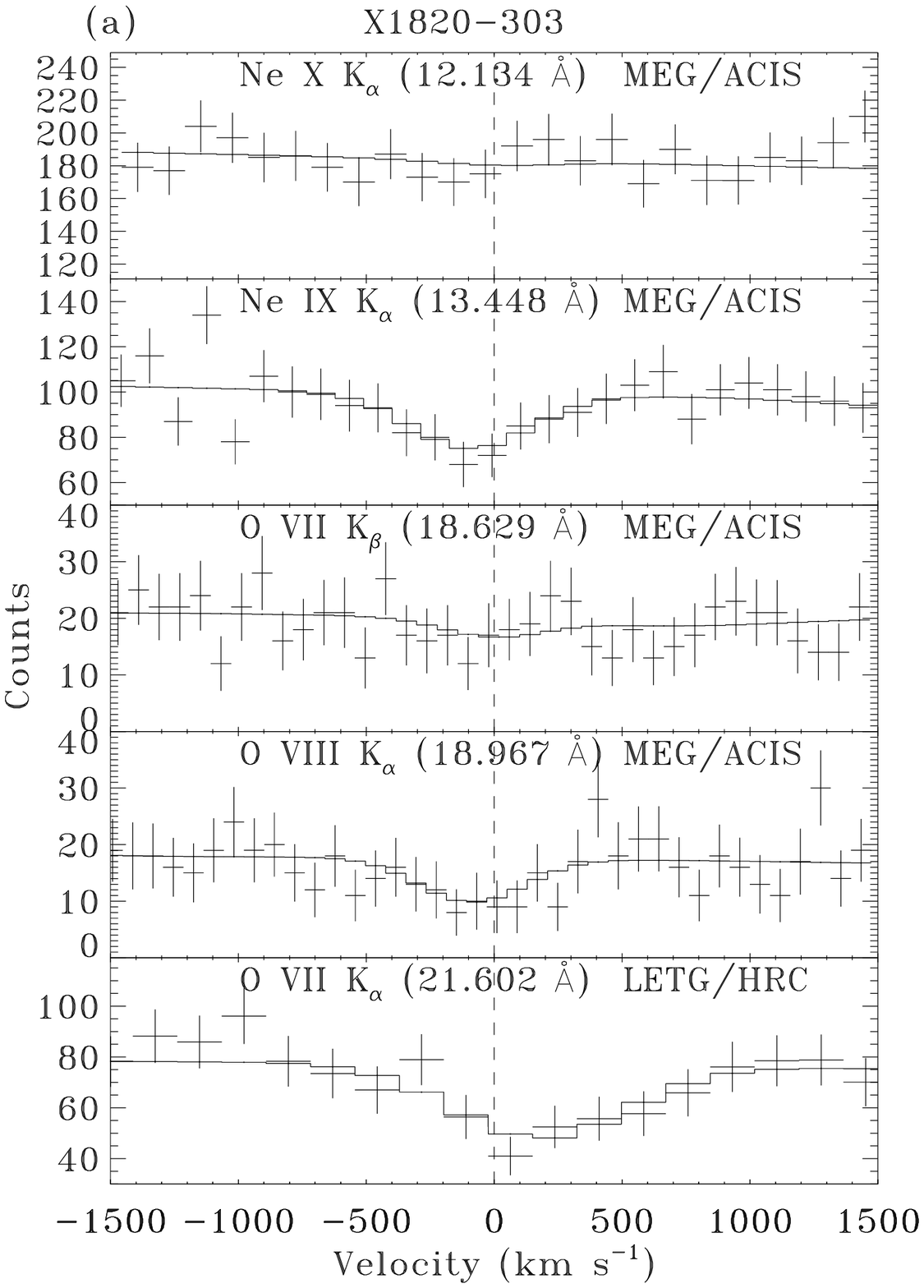,width=0.4\textwidth}
    \psfig{figure=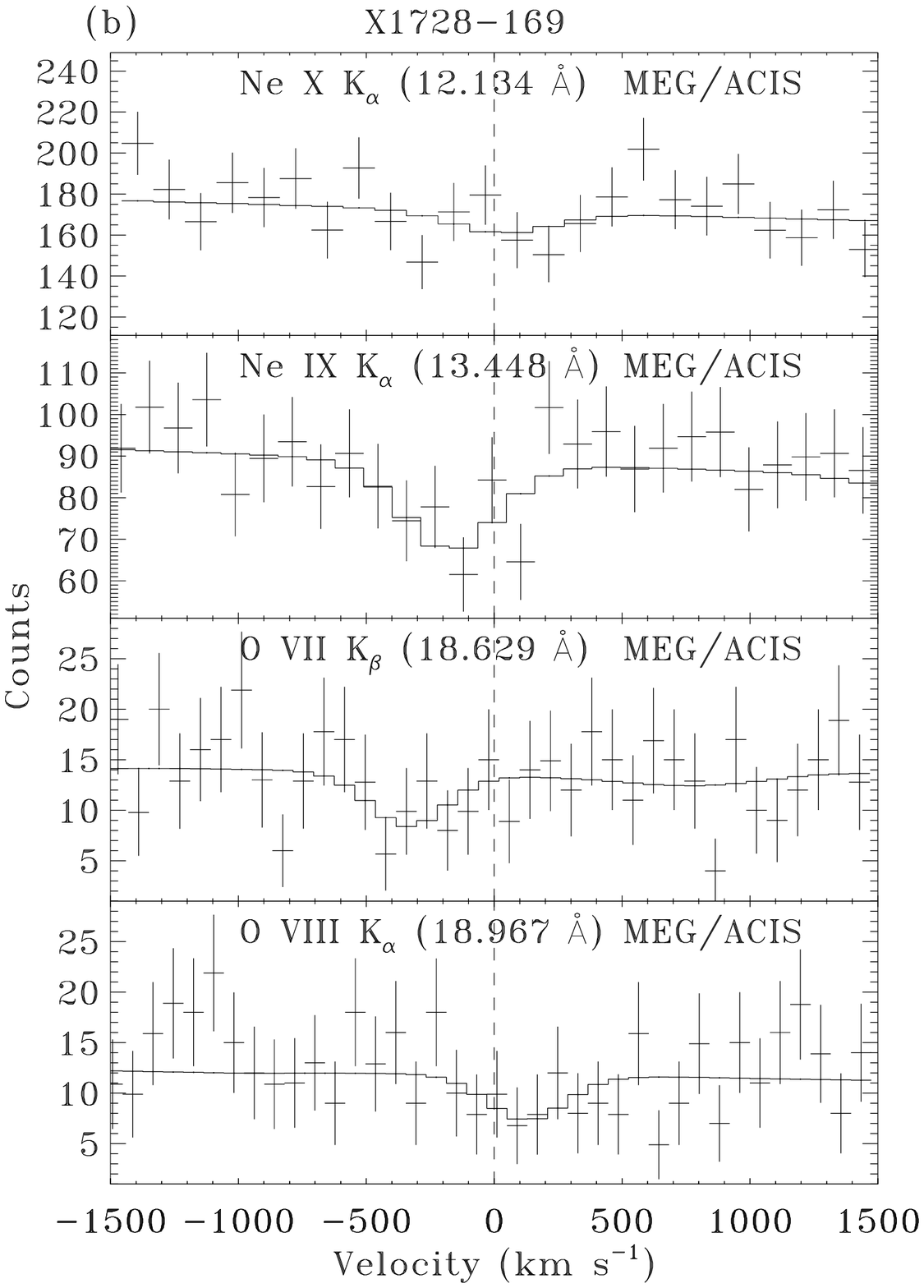,width=0.4\textwidth}
  }}
  \caption{The velocity profiles of the joint-fitted absorption lines.
    The transition and centroid wavelength of each line as well as
    the used instrument are listed 
    in each panel. The {\sl solid} histogram of each panel represents
    the best-fit {\it absline}-modified 
    power-law model folded with the instrumental response.
    The vertical {\sl dashed} line marks the zero velocity.
    A negative velocity indicates a blue shift.
    The bin sizes are the same as in Fig.~\ref{fig:Neon}.
    \label{fig:X1820-1728} }
\end{figure}
\setcounter{figure}{5}
\begin{figure}\centerline{
    \vbox{
      \hbox{
        \psfig{figure=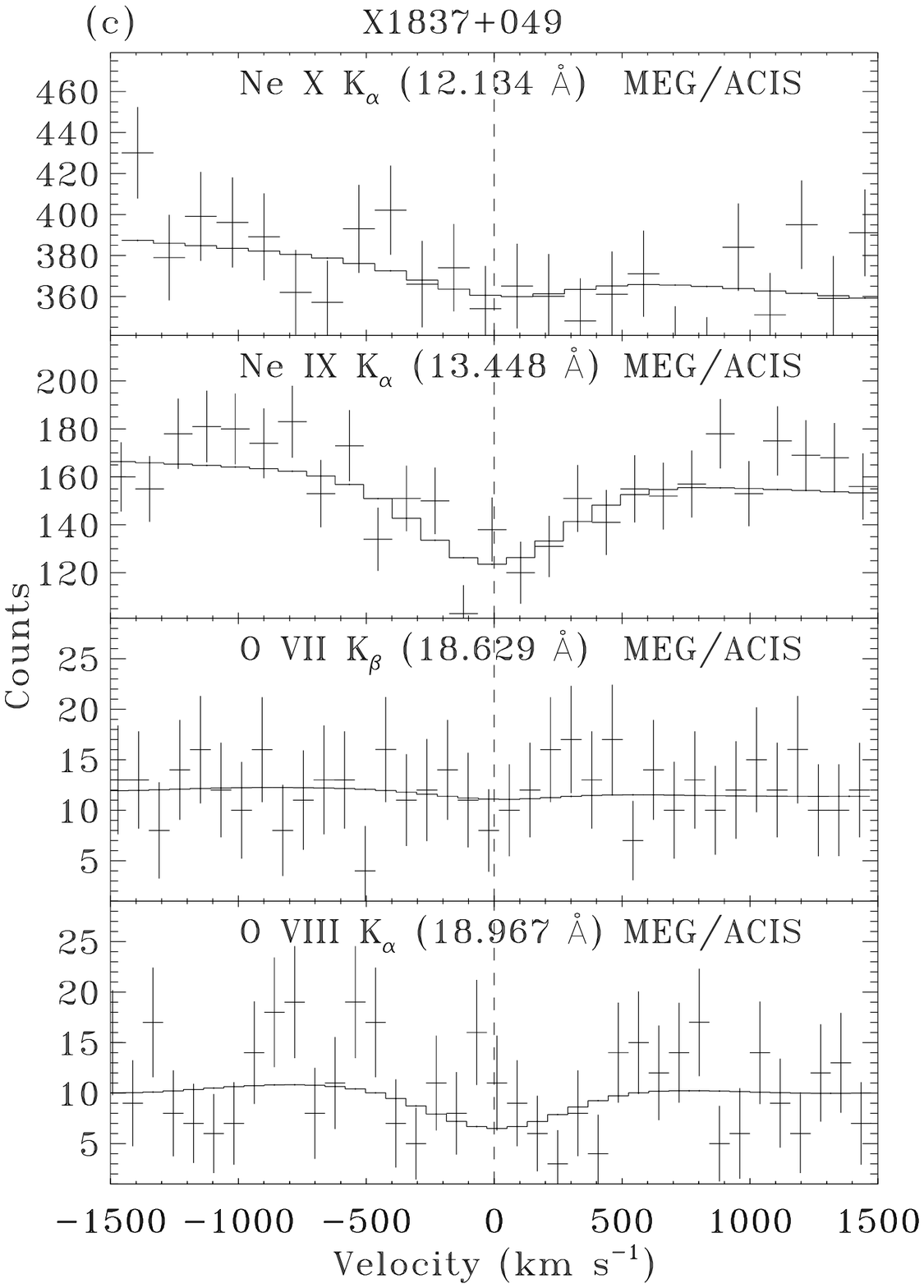,width=0.4\textwidth}
        \psfig{figure=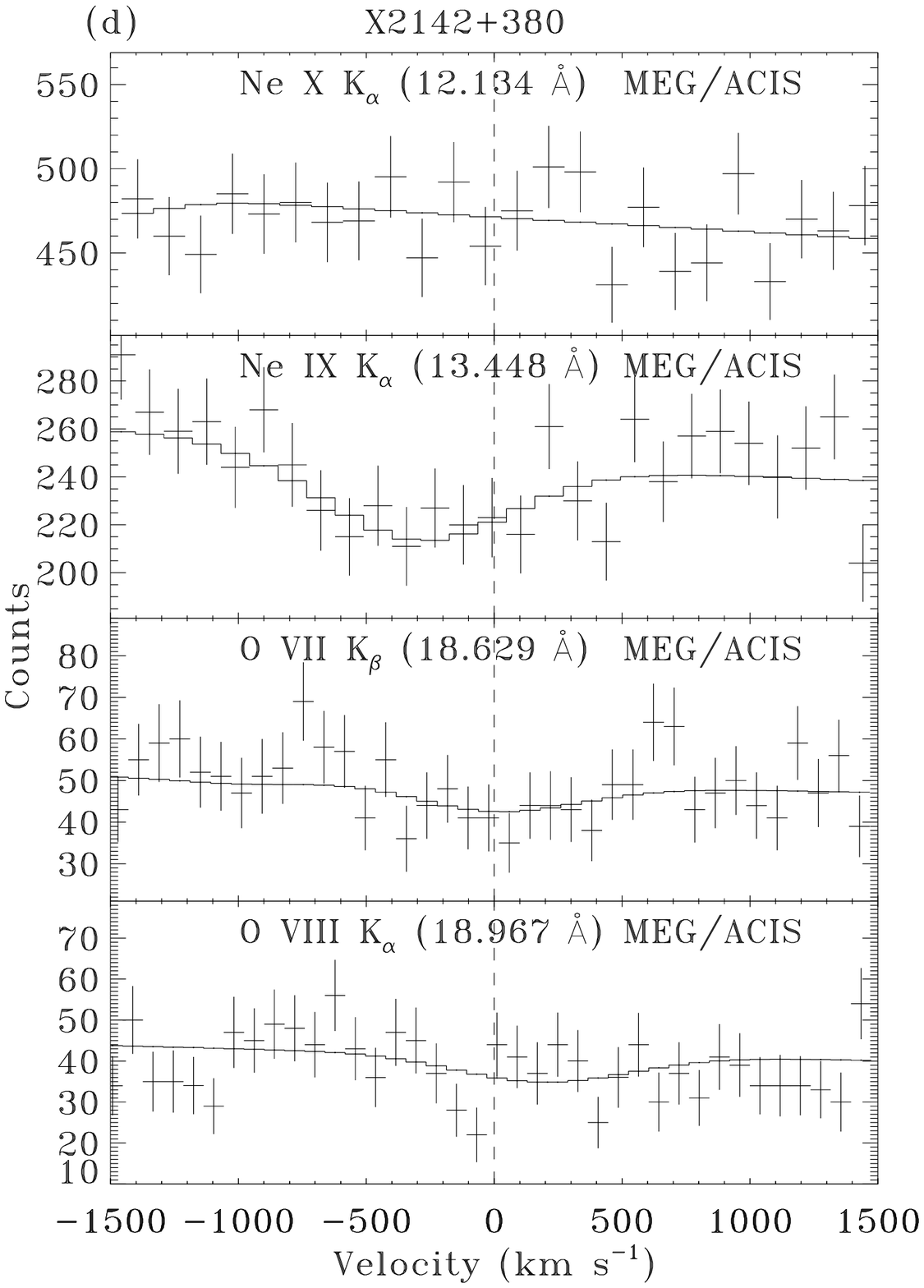,width=0.4\textwidth}
      }\vspace{0.1in}
      \hbox{
        \psfig{figure=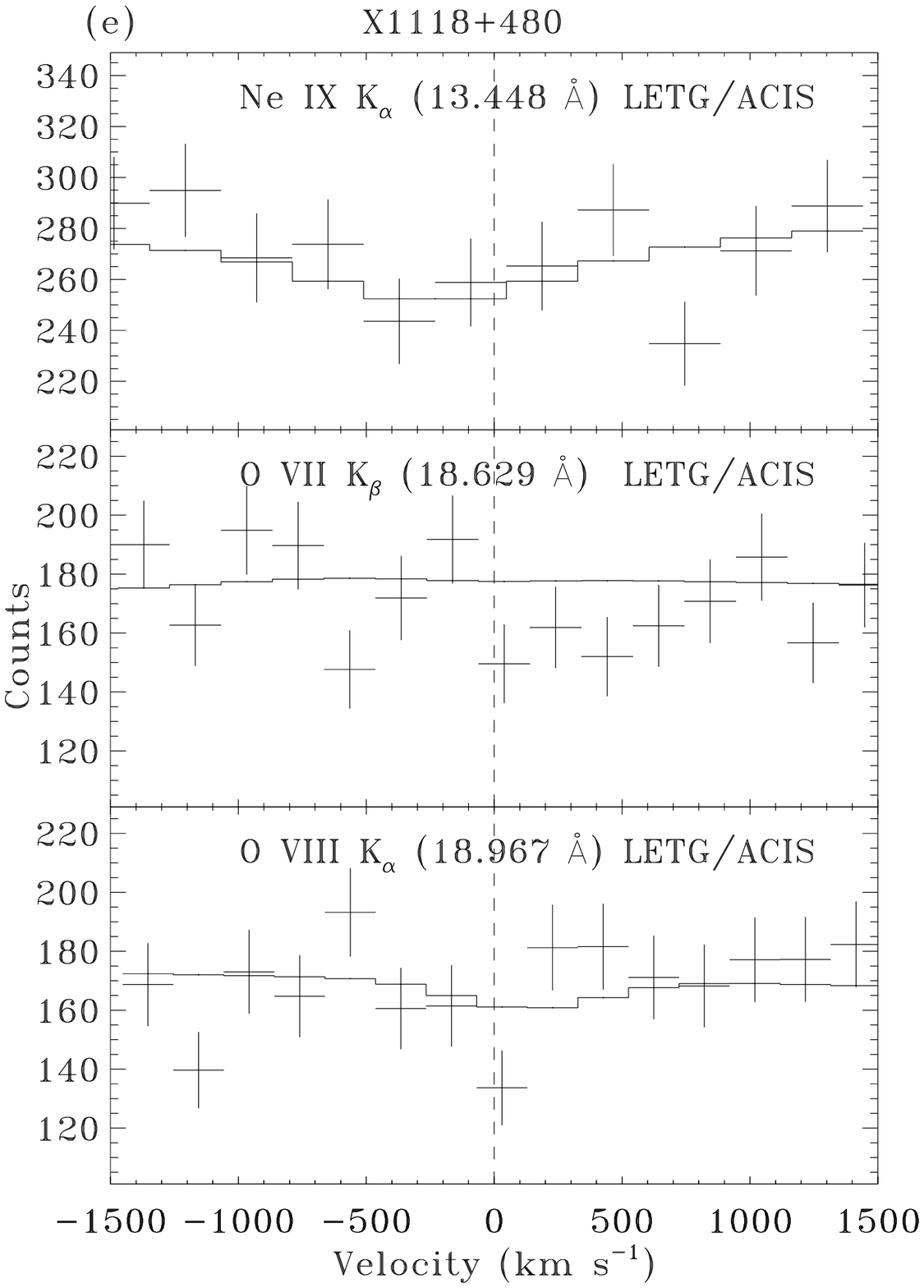,width=0.4\textwidth}
        \psfig{figure=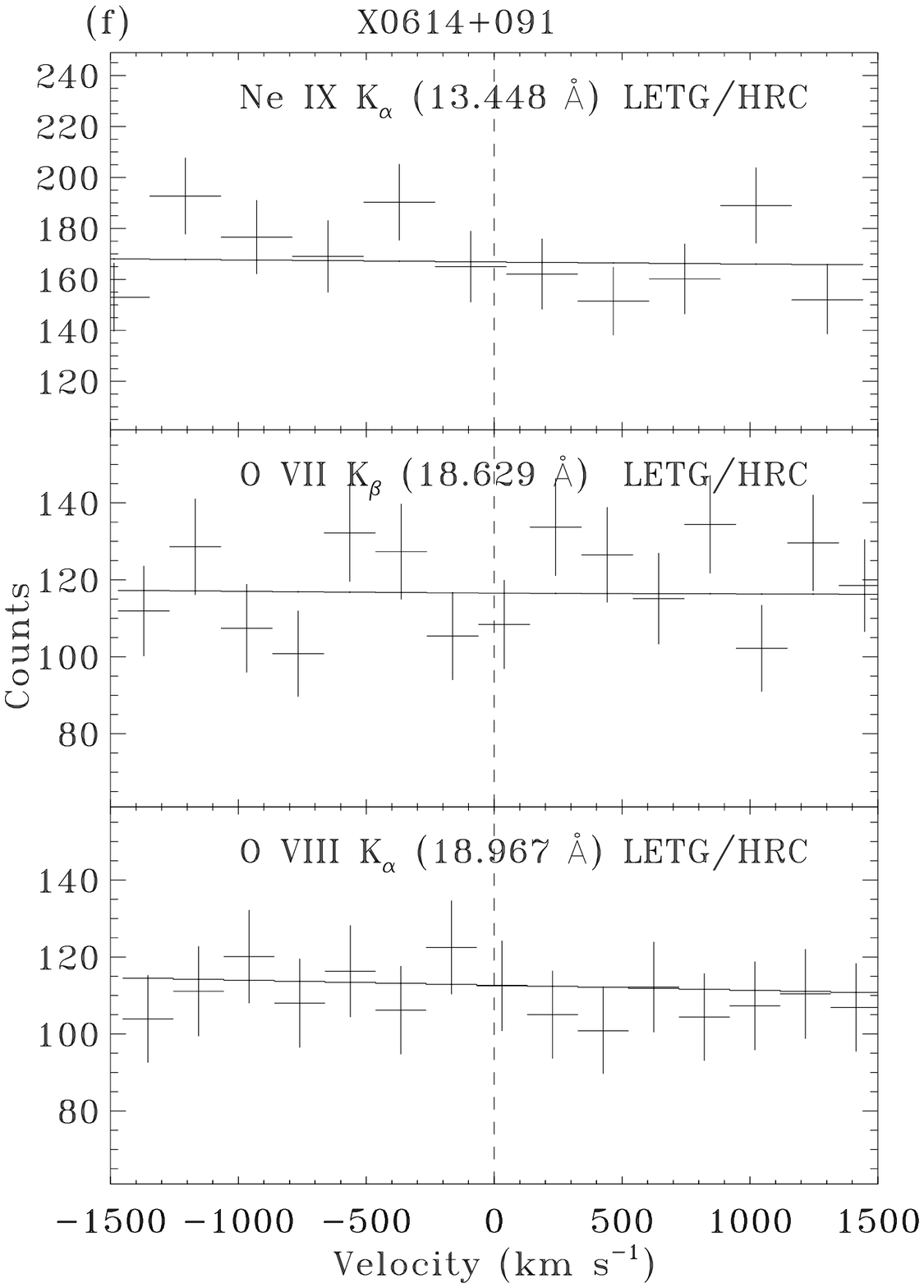,width=0.4\textwidth}
      }
    }
  }
  \caption{Continued. \label{fig:X1837-2142} }
\end{figure}

\setcounter{figure}{5}
\begin{figure}\centerline{
    \vbox{
      \hbox{
        \psfig{figure=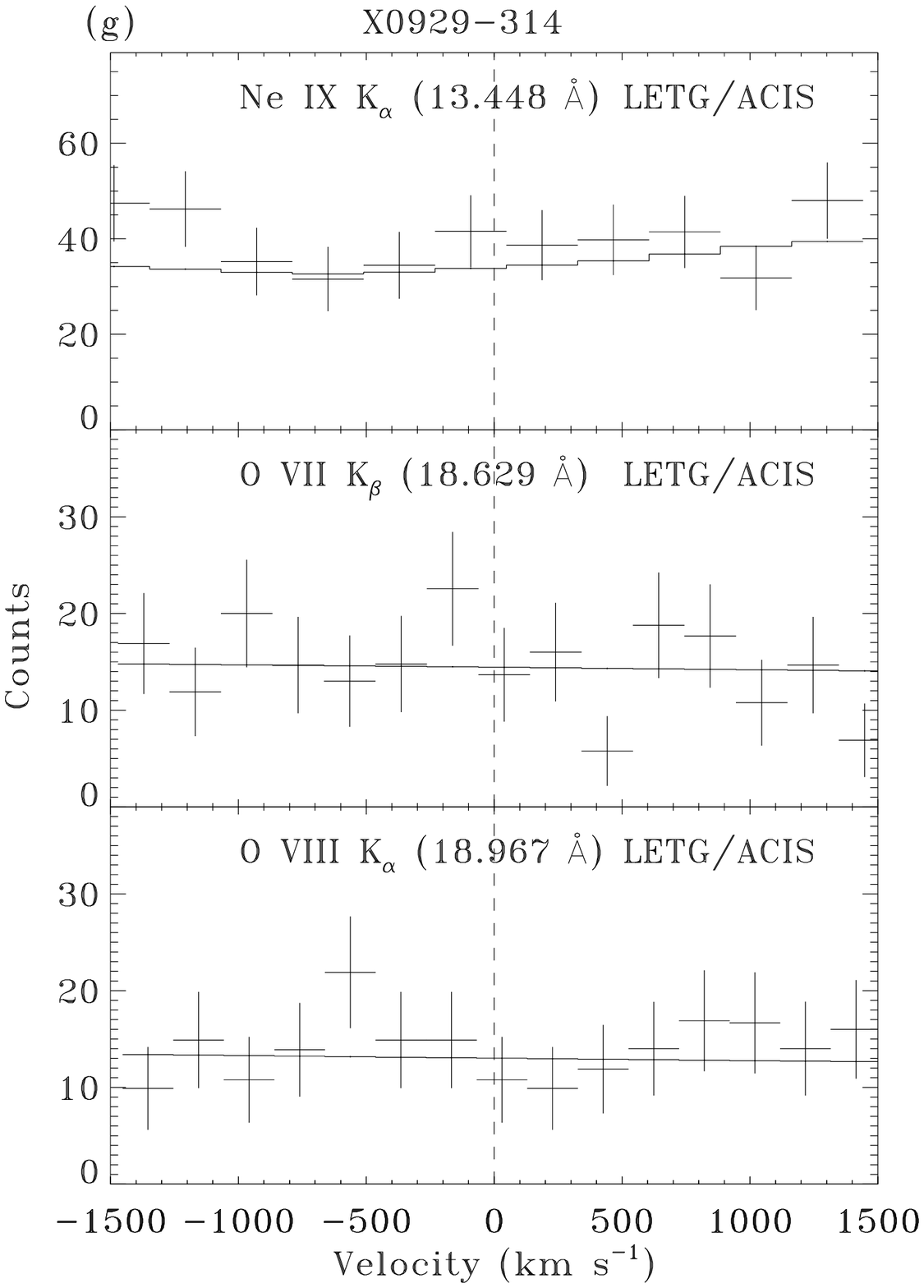,width=0.4\textwidth}
        \psfig{figure=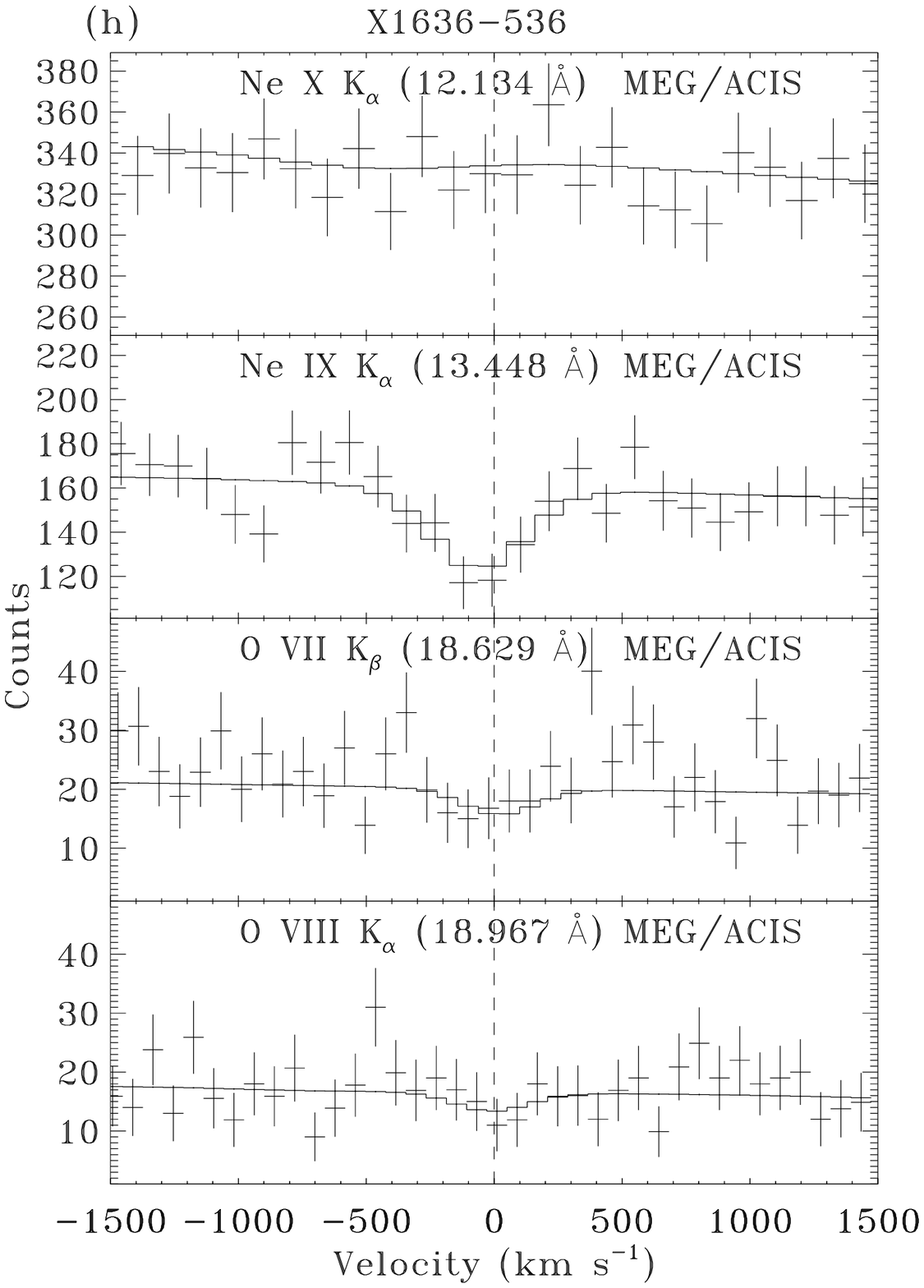,width=0.4\textwidth}
      }\vspace{0.1in}
      \hbox{
        \psfig{figure=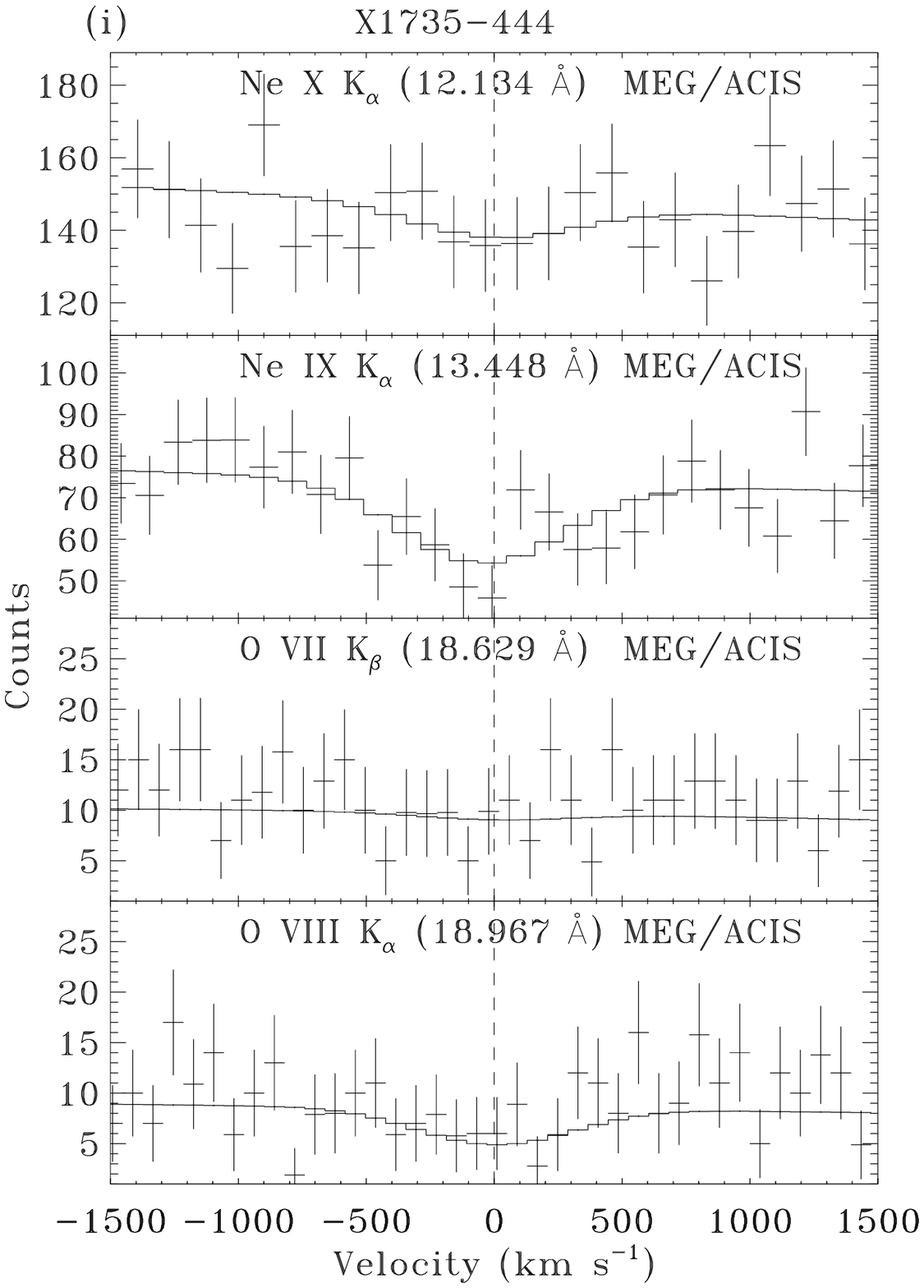,width=0.4\textwidth}
        \psfig{figure=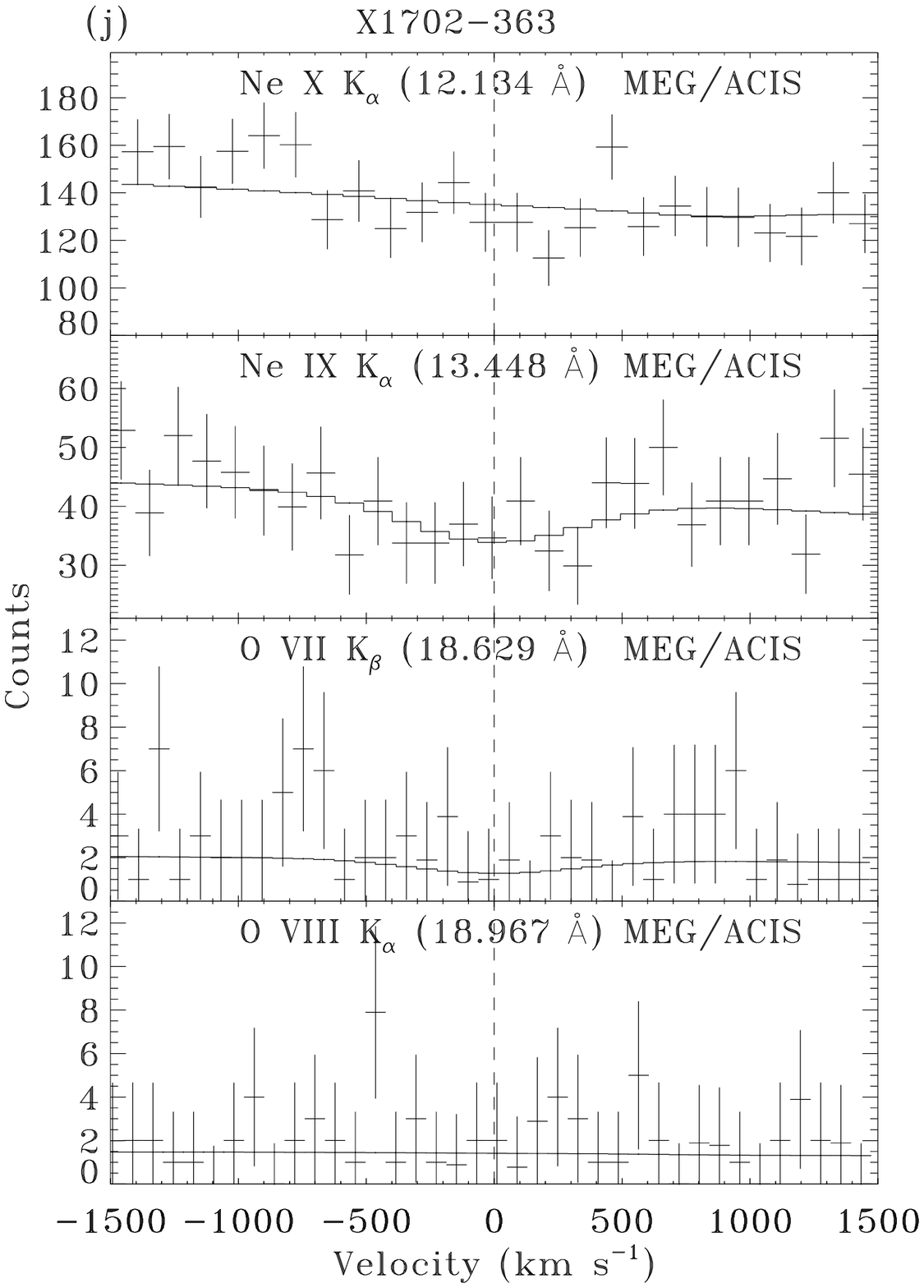,width=0.4\textwidth}
      }
    }}
  \caption{Continued. \label{fig:X0929-1636} }
\end{figure}

Our results are summarized in Table~\ref{tab:fits}.
The sources X1118+480, X0614+091, and X0929--314 are not
included in the table; their poor data quality (in terms 
of both counting statistics and spectral resolution) 
does not give any meaningful upper limits to
either EWs or the ionic column densities.
The $\varepsilon_l$ values and 
their uncertainties constrained from {\sl absline} model 
are consistent with those obtained from the 
{\it Gaussian} model and are thus not repeatedly presented in the table. 
In addition to setting the upper limits to $b_v$ from the 
observed lines, the joint-fits further constrain 
the lower bounds on $b_v$ for four of the sources (Table~\ref{tab:fits}).
If $b_v$ were smaller, the (unresolved) line would then be deeper
in order to provide the same observed absorption. The degree of
saturation $\tau_0$ varies among the jointly-fitted lines 
and therefore affects their relative strengths,
which sets a lower bound on $b_v$. Table~\ref{tab:fits}
further includes a hot gas hydrogen density $n_H$, estimated from
assuming a uniform distribution with a unity filling factor along
each sight-line.

\begin{rotate}
\begin{table}
\caption{Results of {\it Gaussian} and {\it absline} model fits\label{tab:fits}}
\tiny
\begin{tabular}{llc|ccc|cccccc}
\hline
\hline
           &                 &          & \multicolumn{3}{c}{\underline{Gaussian model}} & \multicolumn{6}{c}{\underline{absline model}}\\
           &                 & S/N      & $cz$  & $b_G$  & EW   &       & $b_v$  & log(T) & ${\rm log(N_H)}$ & ${\rm n_H}$ & log(${\rm N_{NeIX}^p}$) \\
Source     & Line            &($\sigma$)& (km/s)& (km/s) & (eV) & Joint & (km/s) & (K)    & (cm$^{-2}$)      & (10$^{-3}$cm$^{-3}$)& (cm$^{-2}$)\\
\hline		  
X1820--303 &  Ne~X~K$_{\alpha}$   &$<1$  & $\cdots$       &$\cdots$& $<0.09$          & $\surd$\\
           &  Ne~IX~K$_{\alpha}$  & 8.6  & -89(-112, 89)   & $<$346 & 0.43(0.28, 0.63) & $\surd$\\
           & O~VII~K$_{\beta}$    &$<1$  & $\cdots$       &$\cdots$& $\cdots$         & $\surd$\\   
           & O~VIII~K$_{\alpha}$  & 6.0  & -86(-201, 152)  & $<$490 & 0.65(0.34, 0.98) & $\surd$\\
           & O~VII~K$_{\alpha}$   & 10.6 & 175(-120, 125)   & $<$347 & 0.58(0.41, 0.79) & $\surd$\\   
           &                     &      &                &        &                  &         & 191(62, 346) & 6.4(6.2, 6.5) & 20.0(19.8, 20.2) & 4.3(2.7, 6.8)   &  $<$18.54 \\  
\hline
X1728--169 &  Ne~X~K$_{\alpha}$   &$<1$  & $\cdots$       &$\cdots$& $<0.36$          & $\surd$\\
           & Ne~IX~K$_{\alpha}$   & 4.9  & -134(-223, 134)  & $<$532 & 0.30(0.14, 0.52) & $\surd$\\
           & O~VII~K$_{\beta}$    & 2.0  & -261(-342, 264)  & $<$375 & 0.39(0.14, 0.65) & $\surd$\\
	   & O~VIII~K$_{\alpha}$  & 2.1  & 178(-298, 617) & $<$531 & 0.43(0.16, 0.70) & $\surd$\\ 
           &                     &      &                &        &                  &         & $<375$       & 6.3(6.0, 6.5) & 19.9(19.7, 22.2) & 5.1(3.2, 1027)   &  $<$18.37 \\  
\hline
X1837+049  &  Ne~X~K$_{\alpha}$   &$<1$  & $\cdots$       &$\cdots$& $<0.11$          & $\surd$\\
           & Ne~IX~K$_{\alpha}$   & 9.9  & 22(-89, 89)   & $<$366 & 0.42(0.37, 0.47) & $\surd$\\ 
           & O~VII~K$_{\beta}$    &$<1$  & $\cdots$	  &$\cdots$&$\cdots$          & $\surd$\\
	   & O~VIII~K$_{\alpha}$  & $<1$ & $\cdots$       &$\cdots$&$\cdots$          & $\surd$\\
           &                     &      &                &        &                  &         & 280(111,443) & 6.45(6.37, 6.54) & 20.1(19.9, 20.2) & 4.8(3.1, 6.1)  &  $<$18.46 \\  
\hline	   
X2142+380  &  Ne~X~K$_{\alpha}$   &$<1$  & $\cdots$       &$\cdots$& $<0.11$          & $\surd$\\
           & Ne~IX~K$_{\alpha}$   & 4.8  & -223(-267, 290) & $<$797 & 0.25(0.14, 0.40) & $\surd$\\
           & O~VII~K$_{\beta}$    & 4.9  & 1(-263, 285)   & $<$369 & 0.24(0.01, 0.41) & $\surd$\\
	   & O~VIII~K$_{\alpha}$  & $<1$ & $\cdots$       &$\cdots$&$\cdots$          & $\surd$\\
	   & O~VII~K$_{\alpha}$   & $<1$ & $\cdots$       &$\cdots$&$\cdots$          &\\
           &                     &      &                &        &                  &         & 383(30, 664) & 6.3(6.2, 6.4) & 19.7(19.5, 19.9) & 2.3(1.4, 3.6)   &  $<$18.10 \\
\hline
X1636--536 &  Ne~X~K$_{\alpha}$   &$<1$  & $\cdots$       &$\cdots$& $<0.09$          & $\surd$\\
           & Ne~IX~K$_{\alpha}$   & 8.4  & -67(-44, 45) & $<$213 & 0.38(0.27, 0.49) & $\surd$\\
	   & O~VII~K$_{\beta}$    & $<2$ & $\cdots$	  &$\cdots$&$\cdots$	      & $\surd$\\
	   & O~VIII~K$_{\alpha}$  & $<2$ & $\cdots$	  &$\cdots$&$\cdots$	      & $\surd$\\
           &                     &      &                &        &                  &         &  $<$105      & 6.3(5.7, 6.8) & 20.2(19.5, 22.3) & 7.9(1.6, 995)   &  $<$18.28 \\
	   
\hline
X1735--444 &  Ne~X~K$_{\alpha}$   &$<1$  & $\cdots$       &$\cdots$& $<0.31$          & $\surd$\\
           & Ne~IX~K$_{\alpha}$   & 9.0  & -22(-156, 200) & $<$724 & 0.78(0.44, 0.84) & $\surd$\\
	   & O~VII~K$_{\beta}$    & $<2$ & $\cdots$	  &$\cdots$&$\cdots$	      & $\surd$\\
	   & O~VIII~K$_{\alpha}$  & 3.8  & -168(-611, 375)&$<$746  & $<$0.90  	      & $\surd$\\
           &                     &      &                &        &                  &         & 369(58, 638) & 6.5(6.3, 6.6) & 20.2(20.0, 20.4) &7.2(4.6, 11)      &  $<$18.84 \\
\hline
X1702--363 &  Ne~X~K$_{\alpha}$   &$<1$  & $\cdots$       &$\cdots$& $<0.28$          & $\surd$\\
           & Ne~IX~K$_{\alpha}$   & 5.5  & -201(-289, 335) & $<$593& 0.50(0.17, 0.60) & $\surd$\\
	   & O~VIII~K$_{\beta}$   & $<1$ & $\cdots$	  &$\cdots$&$\cdots$	      & $\surd$\\
	   & O~VIII~K$_{\alpha}$  &$<1$ & $\cdots$	  &$\cdots$&$\cdots$	      & $\surd$\\
           &                     &      &                &        &                  &         & $<$899       & 6.0(5.6, 7.1) & 19.9(19.4, 20.3) & 5.1(1.6, 13)     & $<$18.62 \\
\hline
\tablecomments{\small The O~VII~K$_\alpha$ line is only searched in the 
  spectra of LETG observations (Table~\ref{tab:observation}). 
  A negative $cz$ value indicates a blue shift.
  In a {\it Gaussian} model fit, $b_G$ is $\sqrt{2}$ times
  the standard deviation (mimicking the velocity dispersion
  $b_v$ in an {\it absline} model fit). The ``Joint'' column 
  marks those absorption
  lines utilized in the joint-fits with the {\it absline} model. 
  The n${\rm{_H}}$ is the averaged
  hot gas density derived from the column density N${\rm{_H}}$ divided 
  by the source distance $D$ listed in Table~\ref{tab:sources}. All
  limits are at the 90\% confidence level. See the text for details.}
\end{tabular}
\end{table}
\end{rotate}

Considering the potential uncertainties in the systematics of the
spectral resolution calibration, we have also estimated the Ne~IX column 
density ${\rm N_{NeIX}^p}$ by fitting individual Ne~IX~K$_\alpha$ lines with 
the natural broadening only (\S\ref{sec:absline}). This ``firm'' 
upper limit to the Ne~IX column 
density is presented in the last column of Table~\ref{tab:fits}.

The joint-fits give direct estimates of $N_H$ and $T$, 
although the uncertainties are large along some sight-lines.
We define a mean measurement $\overline{p}$ and its 90\% upper and lower 
uncertainties $\Delta\overline{p}_\pm$ of a parameter as
\[ \overline{p} = \frac{\sum_{i=1}^7 p_i/\sigma^2_{i}}
{\sum_{i=1}^71/\sigma^2_{i}},~~
\Delta\overline{p}_\pm=\left(\sum_{i=1}^7\frac{1}
	 {\Delta p^2_{\pm i}} \right)^{-1/2},~~{\rm and}~~
\sigma^2_{i} = \Delta p^2_{+i} + \Delta p^2_{-i}, \]
where $p_i$, $\Delta p_{+i}$ and $\Delta p_{-i}$ are the measurement
and its corresponding 90\% upper and lower uncertainties along the $i$th
sight-line with the absorption line(s) detected. We obtain 
$\overline{T}=2.4(2.1, 2.7)\times10^6$ K, 
$\overline{N}_H=8(6, 10)\times10^{19}$ cm$^{-2}$, and
$\overline{n}_H=3.6(3.0, 4.5)\times10^{-3}$ cm$^{-3}$. 

The plasma properties along the sight-line of X1820--303 
are mostly consistent with those obtained by \citet{fut04}
from a curve-of-growth analysis. However, they
prefer a large $b_v$ ($\ge$ 200 km~s$^{-1}$), arguing that otherwise 
the inferred $N_H$ would be comparable to, or even larger than, 
the neutral hydrogen column density observed in the field. 
Our $b_v$ value can be as low as 62 km~s$^{-1}$, and 
the hot gas column density is well below the neutral hydrogen
column density of $1.9\times10^{21}$ cm$^{-2}$ \citep{boh78}. 
This discrepancy is probably due to a significant deviation of 
their used additive {\sl Gaussian} model from the proper 
multiplicative {\sl absline} profile (Fig.~\ref{fig:abs-gau}) and to the
inaccurate error propagation of the temperature in their analysis
(\S\ref{sec:absline}).

\section{Origin of the Hot Gas \label{sec:wind} }

In principle, the X-ray-absorbing hot gas could be local to the binary 
systems. To account for the large column densities estimated above, a plausible
scenario might be the photo-ionized winds from the accretion disks that
are presumably responsible for the X-ray continuous emission. 
Such a scenario, if confirmed, would be interesting in its own right. 
However, we find that it has serious difficulties:
\begin{itemize}
\item An absorption line produced in an accretion disk wind should be 
blue-shifted and broadened with a magnitude comparable to the escape speed
of the accreting compact object (e.g., Ueda \etal 2004). 
The line shift, for example, can be characterized by  
$\Delta\lambda/\lambda_0= v_w/c \sim (2r_g/r_l)^{1/2}$, where $v_w$ is the wind 
velocity, $r_l$ is the starting radius where the absorption
line is produced, $r_g=GM/c^2$ is the gravitational radius. 
Since all the detected X-ray absorption lines are consistent
with $cz\approx0$, using the wavelength accuracy 0.011\AA~of MEG
as $\Delta\lambda$ and assuming the compact object mass $ M=1.4 M_\odot$,
we obtain $r_l\gtrsim9~R_\odot$.
This required $r_l$ value is much greater than the binary separation 
$A$ (Table~\ref{tab:sources}) for all the 
sources with absorption lines detected except for X2142+380.

\item 
The Ne~IX absorption should arise only in a region with
a proper ionization state. If the electron density in the wind can be 
approximated as $n(r)\propto1/r^\alpha$ ($\alpha>1$), 
the ionization parameter can then be written as,
\[ U\equiv L_X/\left[n(r) r^2\right] = 
         L_Xr^{\alpha-2}/\left[N_H(\alpha-1)r_l^{\alpha-1}\right], \]
where $r>r_l$. 
According to \citet{kal82}, Ne~IX is the dominant ionization state of 
neon only when log($U$)$\le 2$ in both optical thin and thick cases. 
We estimate the luminosity of the individual sources to be
in the range of $2-7 \times 10^{37} {\rm ergs~s^{-1}}$ (assuming the source
distances in Table~\ref{tab:sources}). Further assuming
a Ne~IX fraction $\sim$ 0.5 and taking $N_H\sim10^{23}$~cm$^{-2}$ 
(corresponding to the maximum value of the ${\rm N_{NeIX}^p}$; 
Table~\ref{tab:fits}), $\alpha=2$ (constant wind velocity), 
we obtain $r_l\ge7.5~R_\odot$. Again, the required $r_l$
is much larger than the binary separation $A$ 
of all the sources except for X2142+380.

\item 
We should also expect additional signatures of the wind.
If the wind is launched at a radius $r_w$ away from the compact object, the
hot hydrogen column density between $r_w$ and $r_l$ is then
$N_w = N_H\left[(r_l/r_w)^{\alpha-1} -1\right]$, where $N_H$ is  
inferred from the X-ray absorption lines. Taking $r_w=2\times10^8$ cm 
(corresponding to $\sim10^3~r_g, 3\times10^{-3}R_\odot$; Li \& Wang 1999),
$r_l\sim10 R_\odot$, and $\alpha=2$, we got $N_w\sim10^3N_H$. 
Such a wind should then result in strong emission lines, which are, however, 
absent in the spectra of the sources. Furthermore, because 
the ionization state should increase with decreasing radius, 
we should observe a column density
of Ne~X comparable to, or greater than, that of Ne~IX. But we do not detect
any significant Ne~X absorption line in any of the spectra.
\end{itemize}

Based on these arguments, even though not conclusive
(there are other possibilities such as a clumpy wind), 
we find that the X-ray line absorption is unlikely to be 
associated with the LMXBs.  An interstellar origin of the absorption is most 
likely and is assumed in the following discussion.

\section{Spatial Distribution of the Hot Gas \label{sec:spatial} }

To tighten the constraints on the distribution of the
X-ray-absorbing gas, we include the results from the absorption line
study of the X-ray binary LMC X--3 \citep{wang05}
and an AGN MRK~421 \citep{yao05}.
The $N_H$ measurements, as listed in Table~\ref{tab:extragalactic},
are made in the same way as the 7 Galactic LMXBs reported above.
With this set of 9 column density measurements, though still quite sparse, we 
attempt to characterize the overall spatial scale of 
the X-ray-absorbing gas, by assuming two extreme distribution geometry,  
an infinite disk or a sphere.

\begin{deluxetable}{lccc}
\tablewidth{0pt}
\tablecaption{Results of the Sight-lines toward Extragalactic 
  Sources \label{tab:extragalactic}}
\tablehead{
          & ($l, b$)    & D        & ${\rm log(N_H)}$ \\
Source    & (deg)       & (Mpc)    & (cm$^{-2}$)  
}
\startdata
LMC~X--3      & (273.58, -32.08) & 0.05 & 19.7(19.2, 20.5) \\
MRK~421       & (179.83, 65.03)  & 118  & 19.4(19.3, 19.6) 
\enddata
\tablecomments{The $N_H$ values were adopted from 
  \citet{wang05} and \citet{yao05} respectively.
  The distance to MRK~421 was calculated
  by assuming ${\rm H_0=75~km~s^{-1}~Mpc^{-1}}$. } 
\end{deluxetable}

\begin{figure}
  \centerline{
    \hbox{
      \psfig{figure=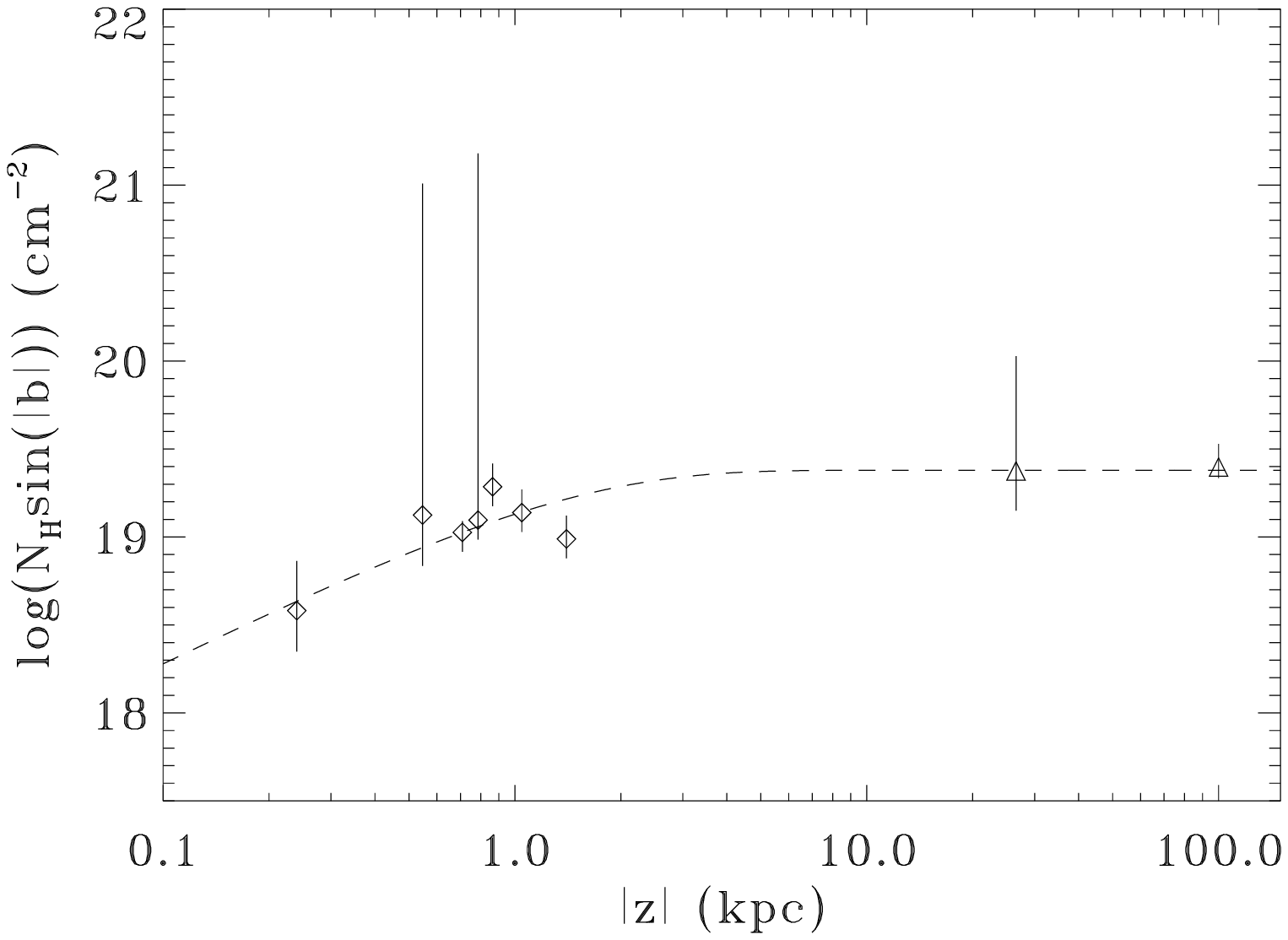,width=0.45\textwidth}
      \psfig{figure=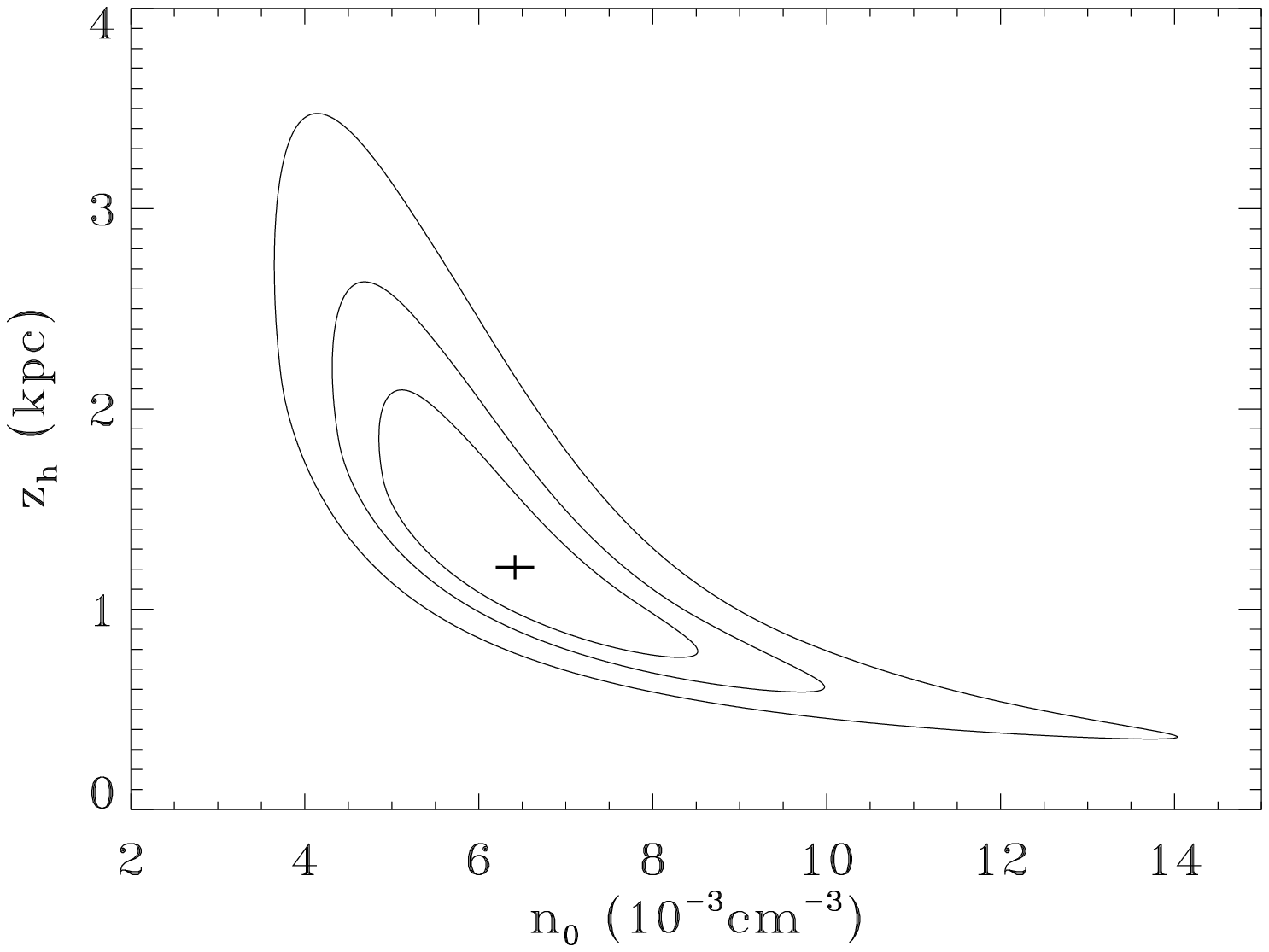,width=0.44\textwidth}
    }}
  \caption{{\sl Left panel}: The vertical hot gas column density as a function 
    of the distance from the Galactic plane. {\sl Diamonds} mark the 
    measurements toward the LMXBs and {\sl triangles} represent the 
    line-of-sights to the extragalactic sources (see the text for details). 
    The {\sl dashed line} is the best-fit exponential function with 
    $n_0 = 6.4\times10^{-3}{\rm cm^{-3}}$ and $z_h = 1.2 {\rm~kpc}$  
    (Eq.~\ref{equ:scale}).  {\sl Right panel}: $z_h$ 
    vs. $n_0$ confidence contours at the 68\%, 90\%, and 99\% confidence levels.
    \label{fig:scale} }
\end{figure}

For the disk distribution, we assume
\begin{equation} \label{equ:expdis}
  n(z) = n_0e^{-|z|/z_h},
\end{equation}
where $n_0$ is the mean gas density at the Galactic plane and 
$z_h$ is the vertical scale height. 
The column density can then be expressed as
\begin{equation} \label{equ:scale}
  N_H=n_0z_h(1-e^{-|z|/z_h})/\sin|b|. 
\end{equation}
A $\chi^2$ fit to the data (Fig.~\ref{fig:scale}) 
gives $n_0=6.4(4.7, 8.8)\times10^{-3}{\rm cm^{-3}}$ 
and $z_h$=1.2(0.7, 2.2) kpc with the best-fit $\chi^2/d.o.f.=7.2/7$.  
Here, we have set the $|z|$ value of MRK~421 
to be $\sim$100 kpc for ease of fitting and plotting; the 
exact value is irrelevant as long as it is much larger than $z_h$. 
Furthermore, the 90\% confidence uncertainties of the individual $N_H$ 
measurements have been scaled by a factor of 0.61 to mimic the 
required 1~$\sigma$ errors in the $\chi^2$ fit. 

\begin{figure}
  \centerline{
    \psfig{figure=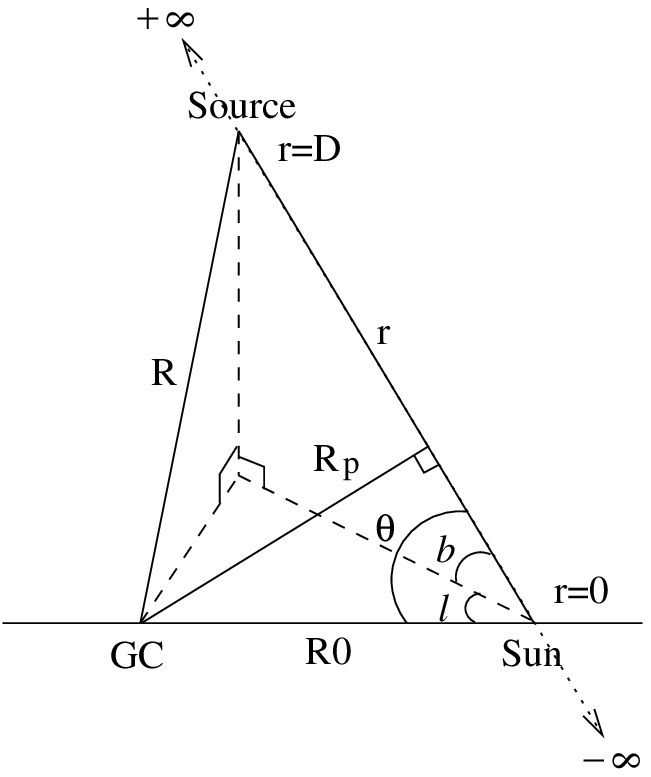,width=0.4\textwidth}
  }
  \caption{Geometry for the Galaxy-centered spherical distribution of the 
    X-ray-absorbing hot gas. \label{fig:sphere_demo}}
\end{figure}

\begin{figure}
  \centerline{
    \hbox{
      \psfig{figure=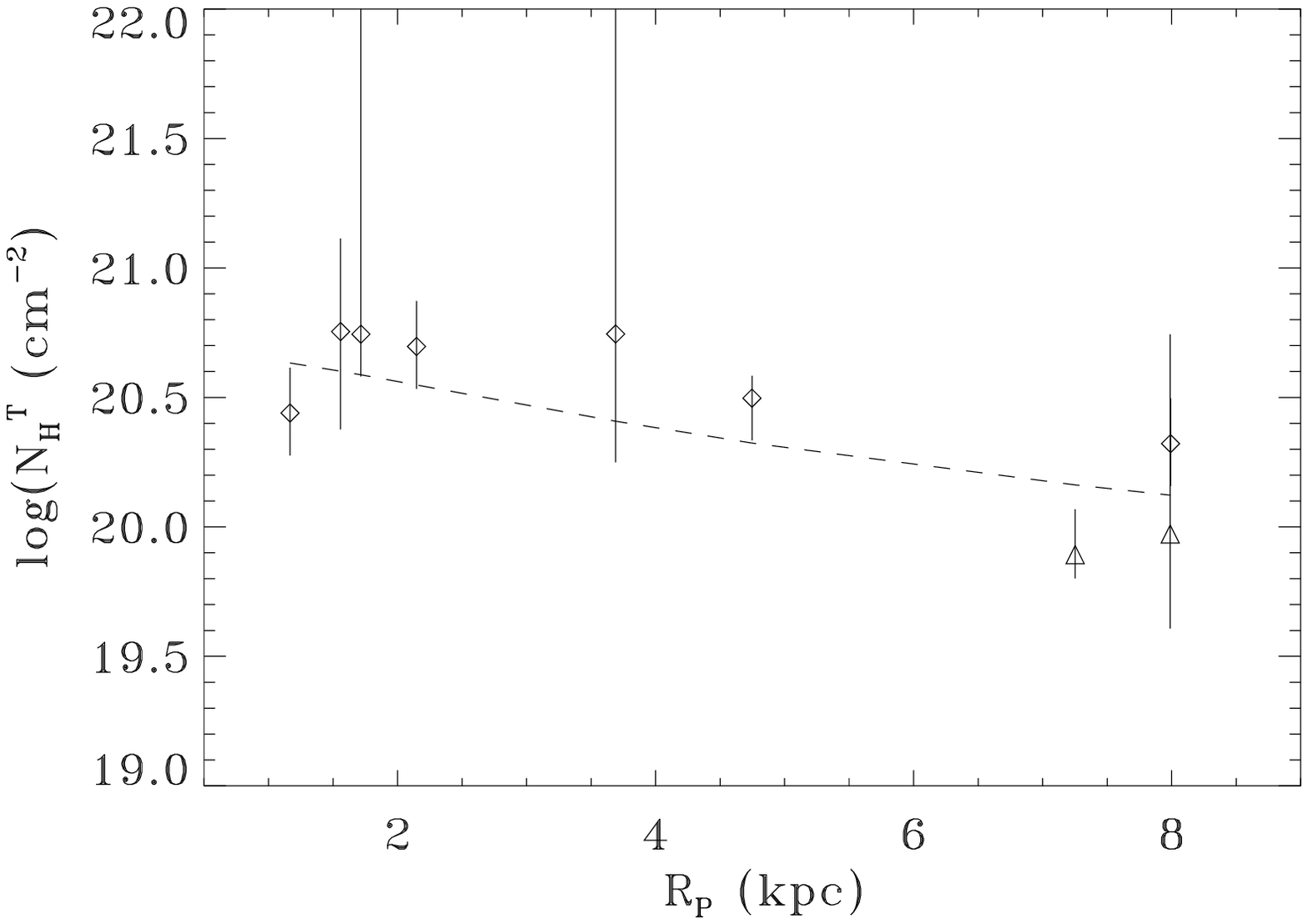,width=0.4925\textwidth}
      \psfig{figure=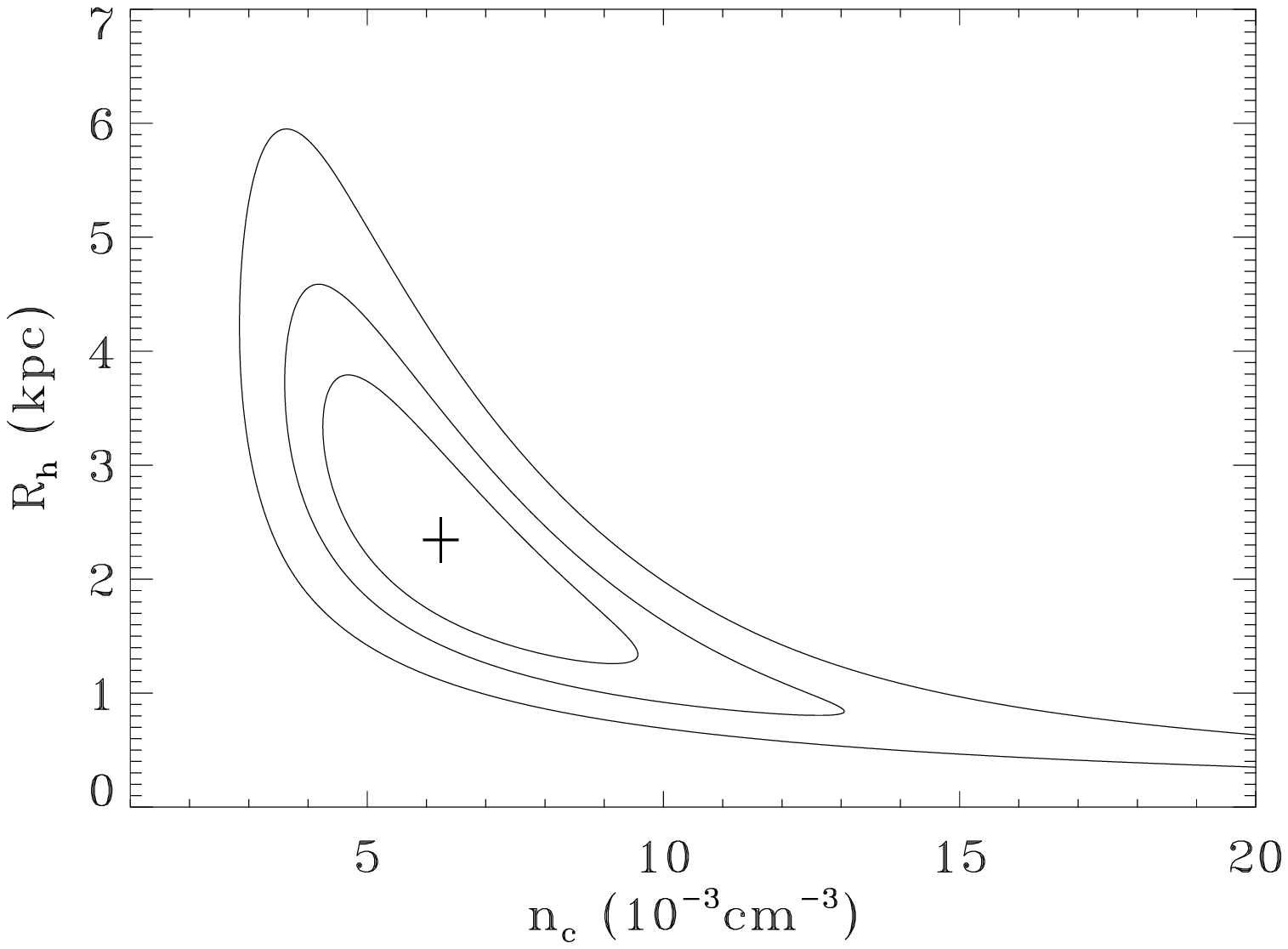,width=0.4675\textwidth}
    }}
  \caption{{\sl Left panel}: The total hot gas column density $N_H^T$ 
    as a function of the off-GC impact radius $R_p$ in the spherical 
    distribution model. $N_H^T$ is calculated
    via an integration along the {\sl dotted line} 
    (i.e., $r = -\infty$ to $\infty$) in Fig.~\ref{fig:sphere_demo}.
    The column density measurements and their errors have been enlarged by
    the ratio of $N_H^T$ to the model $N_H$ between the Sun and the source
    (see Fig.~\ref{fig:sphere_demo}). This plot is for demonstration only; the 
    actual fit is based on the comparison between the measured and modeled 
    $N_H$ values. The meaning of the symbols is the same as that in 
    Fig.~\ref{fig:scale}. The {\sl dashed} line represents the best fit 
    function with $n_c=6.2\times10^{-3}$ cm$^{-3}$, $R_h=2.3$ kpc 
    (Eq.~\ref{equ:sphere_dis}). {\sl Right panel}: $R_h$ vs. $n_c$ contour at
    the 68\%, 90\%, and 99\% confidence levels.
    \label{fig:sphere_scale} }
\end{figure}

For the spherical distribution, we assume
\begin{equation}
n(R) = n_c \left[ 1 + (R/R_h)^2  \right]^{-1},
\end{equation}
adopted from the so-called $\beta$-model 
$n(R) = n_c \left[ 1 + (R/R_h)^2  \right]^{-3\beta/2}$ with $\beta=2/3$
\citep{sar88, jon84},
where $n_c$ is the hot gas density at the Galactic center 
(GC) and $R_h$ is the scale radius. The hot gas column density toward 
a source with Galactic coordinates ($l$, $b$) 
and a distance $D$ (Fig.~\ref{fig:sphere_demo})
can be calculated via the following integration,
\begin{eqnarray}
N_H & = & \int^D_0n_c\left[ 1 + \left(\frac{R}{R_h}\right)^2 \right]^{-1} dr \\
    & = & n_c\int^D_0\left[ 1 + \frac{r^2 + R_0^2 - 2rR_0\cos\theta}{R_h^2} \right]^{-1} dr\\
    & = & n_cR_h\int^{\frac{D-R_0\cos\theta}{R_h}}_{-\frac{R_0\cos\theta}{R_h}}\left( a^2 + x^2 \right)^{-1} dx\\
    & = & \frac{n_cR_h}{a}\left[
      \tan^{-1} \left( \frac{D-R_0\cos\theta}{aR_h} \right) - 
      \tan^{-1} \left( \frac{-R_0\cos\theta}{aR_h} \right)  \right],
    \label{equ:sphere_dis}
\end{eqnarray}
where
$a^2 = 1 + {R_0^2\sin^2\theta}/{R_h^2}$,
$x = ({r - R_0\cos\theta})/{R_h}$,
$R_0$ is the distance between the Sun and GC (taken as 8 kpc in 
this work), and $\theta$ is the angular separation between GC and the source
($\cos\theta=\cos l\cos b$). 
The fit ($\chi^2/d.o.f = 9.4/7$)
of the data with this model (Fig.~\ref{fig:sphere_scale}) is slightly
worse than that with the disk model. The best-fit parameters are 
$n_c=6.2(4.1, 10.0)\times10^{-3}$ cm$^{-3}$, $R_h=2.3(1.2, 3.9)$ kpc. 


With the fitted $n_0$ and $z_h$, or $n_c$ and $R_h$, we estimate 
the total hot gas mass as $\sim2.7\times10^8$ M$_\odot$ and 
$\sim1.2\times10^8$ M$_\odot$ for the disk and the sphere
characterizations within a 15 kpc radius.
This mass estimate is very uncertain as it depends
sensitively on the assumed characterization forms
and their spatial limits, which cannot be constrained
by the existing data.
Taking the CIE radiative cooling function 
${\rm \sim5\times10^{-23}~ergs~cm^3~s^{-1}}$
at $T\sim2.4\times10^6$ K \citep{sut93}, we obtain 
the corresponding cooling rates 
as ${\rm \sim5\times10^{40}~ergs~s^{-1}}$ and
${\rm \sim7\times10^{39}~ergs~s^{-1}}$,
which, within their uncertainties, are comparable to those 
obtained from observations of nearby disk galaxies of 
similar sizes (e.g., Wang \etal 2003). The cooling
probably accounts for only a small fraction of the
expected supernova (SN) mechanical energy input
in the Galaxy (${\rm \sim10^{42}~ergs~s^{-1}}$, assuming
$10^{51}~{\rm ergs}$ per SN and one SN per 30 years).

A more realistic spatial distribution of the hot gas may be between  
the above two extreme characterizations. Also
our model fits are probably quite biased because of the limited number 
of the sight-lines. Nevertheless, the fitted values of $z_h$ and $R_h$
consistently indicate a small scale of
the hot gas distribution. Because the fits incorporate the measurements toward
the two extragalactic sources, this small 
scale suggests that a bulk of the highly-ionized 
X-ray-absorbing gas arises from regions close to the Galactic disk.

In this work, we adopt the element abundances from \citet{wilms00} in which
the abundance ratios of O/H and Ne/H are 0.58 and 0.71 times solar values 
given by \citet{and89}. The real ISM abundances are still very uncertain
(e.g., Savage \& Sembach 1996).
However, if the real abundances are smaller or larger by a factor
than those we adopted, the derived $N_H$ (in \S\ref{sec:results}), 
$n_0$, $n_c$, and the total mass of the hot gas will be larger or 
smaller by the same factor accordingly,
while the inferred $T$ (in \S\ref{sec:results}), $z_h$, and $R_h$ are 
still valid. Since the Ne~IX and the O~VII/O~VIII absorption lines
are used here as the diagnostic of the ionization states and 
the gas temperature (Table~\ref{tab:fits}), 
the results depend on the assumed Ne/O abundance 
ratio. However, this dependence can be partially compensated
by a change in the best-fit $T$ (please refer to 
Fig.~\ref{fig:NeO_T}). Therefore, the $N_H$
measurements should not be very sensitive to the assumed
Ne/O ratio. In general, both $T$ and 
Ne/O could be constrained simultaneously if multiple 
high S/N absorption lines of these two elements are 
available (\S\ref{sec:intro}).

\section{Comparisons with O~VI Absorption and X-ray Emission 
  Measurements \label{sec:OVI} }

We compare the above results from the X-ray absorption lines
with those from observations of Galactic far-UV O~VI absorption lines 
and diffuse soft X-ray emission. Although the O~VI absorption doublet (1031.93
and 1037.62 \AA) has commonly been detected in the spectra
of both extragalactic and Galactic sources, the nature of the
absorbing gas is still very uncertain. 
The O~VI population peaks at $T \sim 3 \times 10^5$ K, near the peak of the 
cooling curve of a CIE plasma. Therefore, 
such intermediate-temperature gas is expected primarily 
at the interfaces between the cool/warm clouds and the hot gas, 
consistent with the finding 
that about half of distinct high-velocity O~VI absorbers have H~I counterparts
\citep{sem03}. The bulk of the observed Galactic O~VI density can be well 
described as a patchy exponential distribution with 
a scale height between 2.3 and 4 kpc (e.g., 
Zsarg\'{o} \etal 2003). This scale height is 
greater than that inferred from our disk modeling of 
the X-ray absorption lines, which may indicate an evolutionary sequence of the 
hot gas. 

The X-ray-absorbing gas should also be responsible for some of the 
O~VI population. We use the constrained $n_0$ and the CIE
ionization fraction of O~VI at $T\sim2.4\times10^6$ K to estimate the O~VI
midplane density as 2.6(1.9, 3.5) $\times10^{-9}$ cm$^{-3}$, which is
$\sim1/7$ of the value obtained from an O~VI survey \citep{jen01}.
Therefore, a significant fraction of the O~VI 
absorption could arise from the X-ray-absorbing hot gas.
Interestingly, the tightly constrained mean
velocity dispersion of the O~VI-absorbing gas \citep{sav03,zsa03}
agrees well with the oxygen thermal velocity ($\sim50$ km~s$^{-1}$) 
at $T\sim2.4\times10^6$ K, but 
is substantially greater than the velocity ($\sim18$ km~s$^{-1}$) at the
O~VI peak temperature. Of course, much of the O~VI line velocity dispersion 
may arise from turbulent and/or differential motion 
(e.g., due to the circular rotation of the disk and halo of the Galaxy).
Detailed cross-correlation between the O~VI and X-ray
absorbing gas along individual sight-lines may provide
insights into the connection between the O~VI- and X-ray-absorbing gases.

Fig.~\ref{fig:galacticmap} shows a {\sl ROSAT} all-sky diffuse 
3/4-keV background map, including the locations of our studied sources
and the two extragalactic sources. 
Among the seven LMXBs with the detected X-ray  
absorption lines (Table~\ref{tab:fits}), five 
(X1820--303, X1728--169, X1636--536, X1735--444, and X1702--363) 
are within a $30^\circ$ radius of the GC;
in this Galactic inner region, the diffuse SXB is greatly 
enhanced in both 3/4 and 1.5 keV bands. 
The bulk of this enhancement may arise from the Galactic 
bulge (e.g., Wang 1998; Almy \etal 2000).
The three sources (X1118+480, X0614+019, and X0929--314),
in which no absorption line is detected, are all toward outer 
Galactic regions with relatively low SXB emission; X1118+480 and
X0614+019 are also very close to us. Our estimated
temperature and density of the X-ray-absorbing hot gas are
between those inferred from the modeling of the X-ray emission gas in 
the solar neighborhood and the Galactic bulge \citep{wang98, kun00}.
These apparent correlation and consistency suggest that the absorption lines 
and the SXB emission arise from the same hot gas. 

\section{Summary and Conclusions \label{sec:sum} }

We have systematically searched for absorption lines produced
by highly ionized species (O~VII, O~VIII, and Ne~IX) in the spectra 
of 10 LMXBs observed with {\sl Chandra} grating instruments.
Much of our analysis is based on the multiplicative absorption line model,
{\it absline}
\footnote{The {\it absline} model, as implemented in XSPEC, can be obtained 
from the authors of this paper.
} 
, which we have constructed to allow for 
both an accurate absorption line profile fit 
and a joint analysis of multiple line transitions from same and/or 
different ions in a CIE plasma.

We detect significant Ne~IX K$_\alpha$ absorption lines
in seven  of the 10 LMXBs.
Three of these seven sources also show evidence for
absorption by O~VII K$_\alpha$, K$_\beta$, and/or O~VIII~K$_\alpha$. 
The detected absorption lines most likely originate in the intervening 
diffuse hot gas, rather than in the LMXBs themselves, because of the 
absence of the expected line centroid shift and broadening as well as
the lack of significant Ne~X absorption and emission lines,
which may be expected from a disk wind.
We estimate the average temperature and the hydrogen density 
of the absorbing gas to be $\sim 2\times10^6$ K and 
$\sim4\times 10^{-3}$ cm$^{-3}$, 
under the assumptions of the CIE, unity filling factor,  
and ISM abundances.

The hot gas is apparently located in and around the Galactic disk and 
preferentially in inner regions
of the Galaxy, consistent with the distribution of the 
diffuse SXB. We have included the column density
measurements inferred from the absorption lines
detected in two extragalactic sources to constrain the global distribution
of the X-ray-absorbing hot gas. Modeled with an infinite disk with
an exponential vertical distribution, the gas has a midplane density
of $\sim6\times10^{-3}$ cm$^{-3}$ and 
a scale height of $\sim1$ kpc. 
The gas, if indeed in a CIE state, can account for $\sim$ 1/7 of 
the O~VI absorption observed in the Galaxy.
Modeled with a spheric radial distribution,
the X-ray-absorbing hot gas has a predicted density of 
$\sim6\times10^{-3}$ cm$^{-3}$ at the GC and a 
core radius of $\sim2$ kpc. But this spherical model 
gives a less satisfactory fit than the disk characterization. 
The small characteristic scales indicate that the bulk, if not all,
of the $cz\sim0$ X-ray-absorbing hot gas detected in extragalactic sources
to date is of Galactic origin. 

While far-UV absorption and X-ray emission studies
have had a long history, X-ray absorption spectroscopy of
the Galactic diffuse hot gas has just begun. 
 Nevertheless,
the results presented above demonstrate the potential of X-ray 
absorption line spectroscopy as a powerful diagnostic tool
for probing the hot gas in our Galaxy. 

\acknowledgments
We thank D. Dewey for his detailed comments on an early version of 
the paper. We are also grateful to J. Kaastro, F. Paerels, T. Tripp, 
and T.-T. Fang for useful discussions and comments on the work, 
which is supported in part by NASA under grants AR4-5004A and 
G04-5046B.

\end{document}